\documentclass[floatfix, letterpaper,english,reprint, secnumarabic, amssymb, nobibnotes, aps, prl,superscriptaddress]{revtex4-2}
\usepackage[T1]{fontenc}
\usepackage[latin9]{inputenc}
\usepackage{babel}
\usepackage{amsmath}
\usepackage{amssymb}
\usepackage{cancel}
\usepackage{graphicx}
\usepackage{wasysym}
\usepackage{color}
\usepackage[unicode=true,pdfusetitle,
 bookmarks=true,bookmarksnumbered=false,bookmarksopen=false,
 breaklinks=false,pdfborder={0 0 1},backref=false,colorlinks=false]
 {hyperref}
 \usepackage{soul}
 
\definecolor{red}{rgb}{1,0,0}

\definecolor{blue}{rgb}{0,0,1}

\definecolor{purple}{rgb}{0.5,0,0.5}
\newcommand{\purple}[1]{{\color{purple} #1}} 

\makeatletter

\pdfpageheight\paperheight
\pdfpagewidth\paperwidth

\makeatother

\begin{document}
\title{Nucleon Electric Dipole Moment from the $\theta$ Term with Lattice Chiral Fermions}
\author{Jian Liang}
\email{jianliang@scnu.edu.cn}
\affiliation{Guangdong Provincial Key Laboratory of Nuclear Science, Institute of Quantum Matter, South China Normal University, Guangzhou 51006, China}
\affiliation{Guangdong-Hong Kong Joint Laboratory of Quantum Matter, Southern Nuclear Science Computing Center, South China Normal University, Guangzhou 51006, China}

\author{Andrei Alexandru}
\affiliation{Department of Physics, The George Washington University, Washington, DC 20052, USA}

\author{Terrence Draper}
\affiliation{Department of Physics and Astronomy, University of Kentucky, Lexington,
KY 40506, USA}

\author{Keh-Fei~Liu}
\affiliation{Department of Physics and Astronomy, University of Kentucky, Lexington,
KY 40506, USA}

\author{Bigeng Wang}
\affiliation{Department of Physics and Astronomy, University of Kentucky, Lexington,
KY 40506, USA}

\author{Gen Wang}
\affiliation{Aix-Marseille Universit\'e, Universit\'e de Toulon, CNRS, CPT, Marseille, France}

\author{Yi-Bo Yang}
\affiliation{CAS Key Laboratory of Theoretical Physics, Institute of Theoretical Physics, Chinese Academy of Sciences, Beijing 100190, China}
\affiliation{School of Fundamental Physics and Mathematical Sciences, Hangzhou Institute for Advanced Study, UCAS, Hangzhou 310024, China}
\affiliation{International Centre for Theoretical Physics Asia-Pacific, Beijing/Hangzhou, China}
\affiliation{University of Chinese Academy of Sciences, School of Physical Sciences, Beijing 100049, China}

\begin{abstract}
We calculate the nucleon electric dipole moment (EDM) from the $\theta$ term with overlap
fermions on three domain wall lattices with different sea pion masses
at lattice spacing 0.11 fm. Due to the chiral symmetry conserved
by the overlap fermions, we have well defined topological charge and chiral limit for the EDM. Thus, the chiral extrapolation can be carried out
reliably at nonzero lattice spacings. We use three to four different partially quenched
valence pion masses for each sea pion mass
and find that the EDM dependence
on the valence and sea pion masses behaves oppositely, which
can be described by partially quenched chiral perturbation theory.
With the help of the cluster decomposition error reduction (CDER)
technique, we determine the neutron and proton EDM at the physical
pion mass to be 
$d_{n}=-0.00148\left(14\right)\left(31\right)\bar\theta$ e$\cdot$fm
and 
$d_{p}=0.0038\left(11\right)\left(8\right)\bar\theta$ e$\cdot$fm.
This work is a clear demonstration
of the advantages
of using chiral fermions 
in the nucleon EDM calculation
and paves the road to future precise studies of the strong $CP$ violation 
effects.
\end{abstract}
\maketitle

\textit{Introduction:}
Symmetries and their breaking are essential topics
in modern physics, among which the discrete symmetries $C$ (charge conjugation),
$P$ (parity), and $T$ (time reversal) are of special importance.
This is partially because the violation of the combined $C$ and $P$
symmetries is one of the three Sakharov conditions~\citep{Sakharov:1967dj}
that are necessary to give rise to the baryon asymmetry of the universe (BAU). 
However, despite the great success of the standard model (SM), the weak
baryogenesis mechanism from the $CP$ violation ($\cancel{CP}$) within
the SM contributes negligibly ($\sim16$ orders of magnitude
smaller than the observed BAU~\citep{Farrar:1993sp,Farrar:1993hn,Gavela:1993ts,Gavela:1994dt,Huet:1994jb}).
This poses a hint that, besides the possible $\theta$ term in QCD, 
there could exist beyond-standard-model (BSM)
sources of $\cancel{CP}$ and thus the study of $\cancel{CP}$ 
plays an important role in the efforts of searching for BSM physics. 

The electric dipole moment of nucleons (NEDM) serves as an important observable to study $\cancel{CP}$. 
The first experimental upper
limit on the neutron EDM (nEDM) was given in 1957~\citep{Smith:1957ht} as $\sim10^{-20}$ e$\cdot$cm.
During the past 60 years of experiments,
this upper limit has been improved by 6 orders of magnitude. The most recent experimental
result of the 
nEDM is $0.0(1.1)(0.2)\times10^{-26}$ e$\cdot$cm~\citep{Abel:2020gbr}, 
which is still around 5 orders of magnitude
larger than the contribution that can be offered by the weak $\cancel{CP}$ phase. 
Currently, several experiments are aiming at improving the
limit down to $10^{-28}$ e$\cdot$cm in the next $\sim$10 years.
This still leaves plenty of room for the study of $\cancel{CP}$ from BSM interactions and the QCD $\theta$ term.

As a reliable nonperturbative method for solving the strong interaction,
lattice QCD provides us the possibility of studying the nucleon EDM (NEDM) from
first principles and with both the statistical and systematic
uncertainties under control. To be specific, lattice QCD can be used
to calculate the ratio
between the neutron and proton EDM induced by strong $\cancel{CP}$ and the parameter
$\bar{{\theta}}$, which is the most crucial theoretical input to determine $\bar{{\theta}}$ from experiments.

Many lattice calculations have been carried out on this topic. However,
there was a watershed in 2017 when it was pointed out~\citep{Abramczyk:2017oxr}
that all the previous lattice calculations, e.g.~\citep{Shintani:2005xg,Berruto:2005hg,Guo:2015tla,Shintani:2015vsx,Alexandrou:2015spa},
used a wrongly defined $\cancel{CP}$ form factor
such that all of those old results need a correction. 
Although the fixing is numerically straight forward, 
none of the previous lattice calculations gives statistically
significant results after the fixing, leaving a great challenge
to the lattice community.
Since then, several
attempts~\citep{Syritsyn:2019vvt,Dragos:2019oxn,Alexandrou:2020mds,Bhattacharya:2021lol} 
have been made to tackle the problem, 
but the signal-to-noise ratios of the new results are 
still not satisfying, and
no calculation performed directly at the physical point
gives nonzero results.

A possibility
to bypass this difficulty is to perform the computations with several
heavier pion masses and extrapolate to the
physical point. However,
only with chiral fermions can a correct chiral limit be reached at
finite lattice spacings. Otherwise, extrapolating to the continuum
limit for each pion mass becomes an inevitable prior step before a reliable
chiral extrapolation, which complicates the calculation and 
potentially leads to hard-to-control systematic uncertainties.
The best result, so far, of this approach, using clover fermions, obtained a 2-sigma signal~\cite{Dragos:2019oxn}.

In this article, we demonstrate that 
using chiral fermions 
to extrapolate to the physical point from heavier pion masses is
the most efficient choice to study NEDM on the lattice
at the current stage.
We employ 3 gauge ensembles with different sea pion masses
ranging from $\sim$300 to $\sim$600 MeV and we use 3 to 4 valence
pion masses on each lattice. 
Therefore, we can study both the valence
and sea pion mass dependence of the NEDM and better control the chiral
extrapolation.
The results we obtain at the physical pion mass are
$d_{n}=-0.00148\left(14\right)\left(31\right)\bar\theta$ e$\cdot$fm
and 
$d_{p}=0.0038\left(11\right)\left(8\right)\bar\theta$ e$\cdot$fm
for neutron and proton, respectively.

\textit{Nucleon EDM and the $\theta$ term:}
The QCD Lagrangian in Euclidean space with the $\theta$ term reads (detailed conventions can be found in the Supplemental Materials~\cite{supplemental}):
\begin{equation}
{\cal L}^{E}=\bar{\psi}\left(D\!\!\!\!/^{E}+m_q\right)\psi+\frac{1}{2}\mathrm{Tr}
[F_{\mu\nu}^{E}F^{E,\mu\nu}-{i}\bar{\theta}\frac{g^{2}}{8\pi^{2}}F_{\mu\nu}^{E}\tilde{F}^{E,\mu\nu}],
\label{Lagrangian}
\end{equation}
where $\tilde{F}^{E,\mu\nu}=\epsilon^{\mu\nu\rho\sigma}F_{\rho\sigma}^E$.
The effective parameter $\bar{\theta}=\theta+\frac{1}{N_{f}}{\rm Arg}{\rm Det}\left[M\right]$
where $\theta$ is the original coefficient of the $\theta$ term
and $M$ is the quark mass matrix generated by the spontaneous breaking
of $SU(2)\times U(1)$ in the electroweak sector.  
For simplicity,
we will not distinguish $\theta$ and $\bar{\theta}$ in
the following content. 
A crucial point is that, if 
${\rm Det}\left[M\right]=0$, phase of the $U_{A}(1)$ transformation
is arbitrary, which means one can always find a chiral rotation that lets
$\bar{\theta}=0$, leaving no net effect of $\cancel{CP}$. This indicates
a zero NEDM in the chiral limit~\cite{Baluni:1978rf}, which poses a very strong constraint in the
chiral extrapolation numerically. However, 
as mentioned before, for lattice fermions which violate the chiral symmetry 
this constraint cannot be used at finite lattice spacing.

Given that $\theta$ is small, 
one can expand the theta term in the action in the path integral
and obtain the correlation functions and matrix elements to the leading order in $\theta$ 
as $_{\theta}\langle...\rangle_{\theta}=\langle...\rangle+{i}\theta\langle...Q_{t}\rangle$, 
where $|0\rangle_\theta$ denotes the vacuum with the $\theta$ term (namely, the $\theta$ vacuum),
and $Q_{t}=\int d^{4}xq_{t}(x)\equiv\frac{g^{2}}{16\pi^{2}}\int d^{4}x {\rm Tr}\left[F_{\mu\nu}^{E}(x)\tilde{F}^{E,\mu\nu}(x)\right]$
is the topological charge of the gauge field geometrically.
Based on this expansion, 
the $\cancel{CP}$
electromagnetic (EM) form factor $F_{3}(q^2)$
can be extracted from
normal and $Q_t$ weighted nucleon matrix elements 
with initial momentum $p_i=(m,\vec{0})$ and final momentum $p_f=(E_f,\vec{q})$ as~\cite{supplemental}
\begin{align}
F_{3}(q^2)&=\frac{2m}{E_f+m}\left\{\frac{2E_f}{q_i}\frac{{\rm Tr}
\left[\Gamma_{i}M_{4}^{(3)Q}\right]}{{\rm Tr}\left[\Gamma_{e}M^{(2)}\right]}-\alpha^1G_{E}(q^2)\right\},\nonumber\\
G_{E}(q^2)&=\frac{2E_f}{E_f+m}\frac{{\rm Tr}
\left[\Gamma_{e}M^{(3)}_{4}\right]}{{\rm Tr}\left[\Gamma_{e}M^{(2)}\right]},\ \alpha^1=\frac{{\rm Tr}\left[\gamma_{5}M^{(2)Q}\right]}{2{\rm Tr}\left[\Gamma_{e}M^{(2)}\right]},\label{eq:R3}
\end{align}
where the matrix elements are
\begin{align}
M^{(2)}&=\langle N(p_f)|N(p_i)\rangle, \nonumber\\
M^{(3)}_{\mu}&=\langle N(p_f)|V_{\mu}(0)|N(p_i)\rangle,\nonumber\\
M^{(2)Q}&=\langle N(p_f)|Q_t|N(p_i)\rangle,\nonumber\\
M^{(3)Q}_{\mu}&=\langle N(p_f)|Q_tV_{\mu}(0)|N(p_i)\rangle,
\end{align}
with $V_{\mu}$ being the EM current operator,
$\Gamma_{e}=\frac{1+\gamma_{4}}{2}$ is the unpolarized spin projector, 
$\Gamma_i={-i}\gamma_5\gamma_i\Gamma_{e}$ the polarized projector along the $i$'th direction,
$q^2=(p_f-p_i)^2=-Q^2$ the momentum transfer,
and $q_i$ the nonzero component of the momentum transfer.
The above formalism is the same for both neutron and proton.
In the end, the nucleon EDM can be extracted from the $\cancel{CP}$
form factor $F_{3}(q^2)$ in the forward limit for neutron and proton respectively using
\begin{equation}
d_{n/p}=\frac{F_{3,{n/p}}\left(q^{2}\to0\right)}{2m}\,\,\theta.
\end{equation}

An interesting fact, as seen in Eq.~(\ref{eq:R3}),
is that the neutron $\cancel{CP}$ form factor at the zero momentum transfer limit, $F_{3,n}(0)$ has no $\cancel{CP}$ angle $\alpha^{1}$ dependence since
$G_{E,n}(0)=0$, and thus one actually needs no information
about $M^{(2)Q}$ in the neutron case.

\begin{table}
\centering{}\caption{Parameters of the RBC/UKQCD ensembles: 
label, sea and valence pion masses, and the number of configurations. \label{tab:Parameters-of-the}}
\begin{tabular}{c|c|cccc|c}
\hline 
label &  $m_{\pi,s}$ (MeV) &\multicolumn{4}{c|}{$m_{\pi,v}$ (MeV)} & $N_{{\rm cfg}}$\\
\hline 
\hline 
24I005 & 339  & 282 & 321 & 348 & 389 & 805\\
24I010 & 432  & &  426 & 519 & 600 & 508\\
24I020 & 560  &  & 432 & 525 & 606 & 552\\
\hline 
\end{tabular}
\end{table}

\textit{Numerical setups:}
This study is carried out on three $2+1$-flavor RBC/UKQCD gauge ensembles of domain wall fermions~\citep{Aoki:2010dy} 
with the same lattice spacing 0.1105(3) fm and lattice volume $24^3\times 64$ but different sea quark masses. 
Using the overlap fermion action~\citep{Neuberger:1997fp} on the HYP (hyper-cubic) smeared~\cite{Hasenfratz:2002rp} gauge links,
multiple partially quenched valence
quark masses (as listed in Table~\ref{tab:Parameters-of-the} with other parameters) 
are calculated utilizing the multi-mass inversion algorithm; 
thus both the sea and valence pion mass dependencies of NEDM can be studied
and the chiral
extrapolation can be more reliable.

Generally, using overlap fermions can be ${\cal O}(100)$ times more costly 
compared to the traditional Wilson-like discretized fermion actions.
To improve the computational efficiency, 
12-12-12 grid sources with $Z_{3}$-noise and Gaussian smearing are
placed at $t_{{\rm {src}}}=0$ and $t_{{\rm {src}}}=32$
in one inversion with randomly chosen spatial positions on different
configurations,
and low-mode substitution (LMS)~\citep{Li:2010pw} is applied to suppress the statistical contamination between different source positions. 
We also use the stochastic sandwich method (SSM)~\citep{Yang:2015zja} with LMS 
to make the cost of using multiple nucleon sinks be additive instead of multiplicative. 
We use 8 sets of source noises and 16 sets of sink noises (for each of the source-sink separations $6a$, $7a$, and $8a$) 
to improve the statistics. 
Five nonzero momentum transfers are calculated such that we can reliably do the $q^{2}$ extrapolation to get $F_{3}\left(0\right)$;
the details of the $q^2$ extrapolation are given in the Supplemental Materials~\cite{supplemental}.

\begin{figure}
\begin{centering}
\includegraphics[scale=0.48,page=1,width=0.22\textwidth]{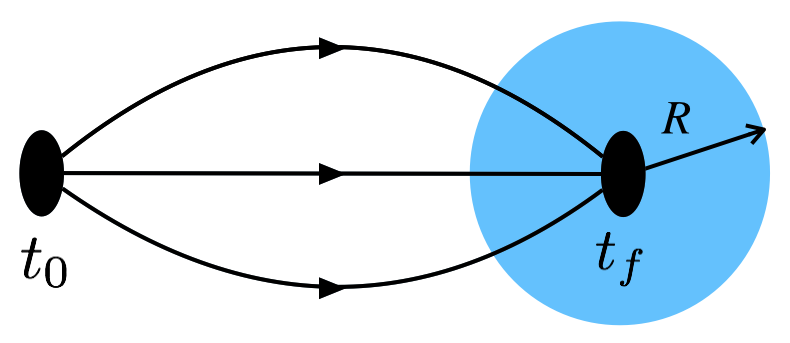}
\includegraphics[scale=0.48,page=1,width=0.22\textwidth]{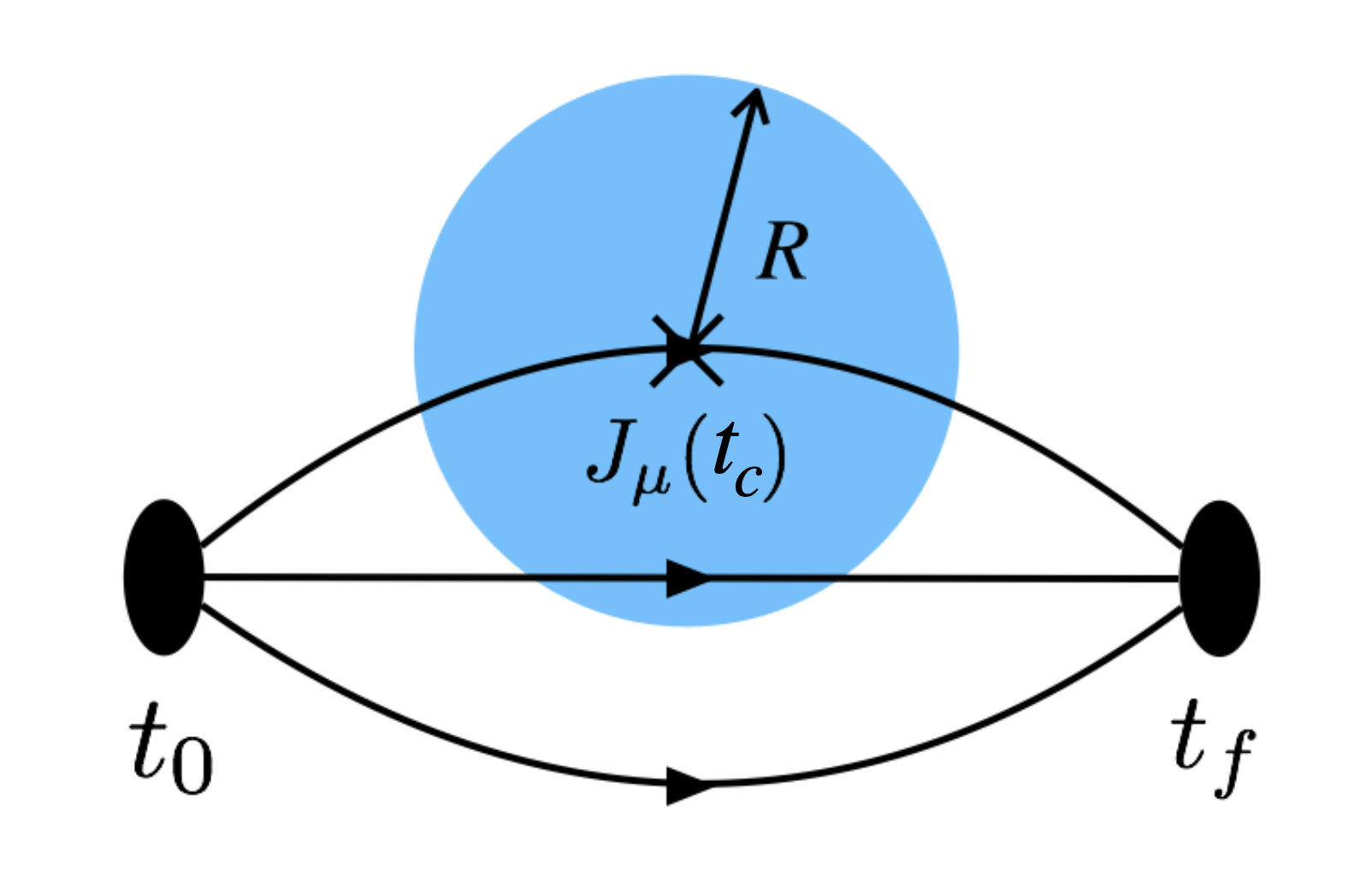}
\par\end{centering}
\centering{}\caption{Illustration of the CDER technique used when computing the correlation
functions with the local topological charge summed inside the sphere with radius $R$. 
\label{fig:Illustration-of-the}}
\end{figure}

\textit{CDER improvement and results:}
To further suppress the statistical uncertainty of $M^{(2)Q}$ and $M^{(3)Q}$, 
we take advantage of 
a technique called cluster decomposition error reduction (CDER) for the disconnected insertion~\cite{Liu:2017man}. 
As illustrated in Fig.~\ref{fig:Illustration-of-the}, we write the total topological charge as the summation of the local charge density $q_t(x)$
derived from the overlap operator~\cite{Adams:1998eg,Fujikawa:1998if} as
$q_{t}(x)=\frac{1}{2}{\rm Tr}\left[\gamma_{5}D_{{\rm ov}}(x,x)\right]$, 
where the trace is over the color-spin indices,
and convert the two-point function weighted with the total topological charge $Q_t$ 
into a summation of the three-point functions involving $q_t(x)$
\begin{align}
G^{(2)Q}&=\sum_{\vec{x}}\left\langle\sum_{r}q_t\left(x+r\right)\chi\left(x\right)\bar{\chi}\left(t_{0},{\cal G}\right)\right\rangle,
\end{align}
where $\chi$ is the nucleon interpolating operator, ${\cal G}$
denotes the source grid, and $x=\left(t_{f},\vec{x}\right)$.
We then use the cluster decomposition property to limit the sum to a range commensurate with the correlation length
\begin{align}
G^{(2)Q}&\sim\sum_{\vec{x}}\left\langle\sum_{r}^{|r|<R}q_{t}\left(x+r\right)\chi\left(x\right)\bar{\chi}\left(t_{0},{\cal G}\right)\right\rangle\nonumber\\
&\sim M^{(2)Q}+{\cal O}(e^{-\delta mt_f},e^{-m_{\eta}R}),\label{eq:CDER-2pt}
\end{align}
which reduces the variance by a volume factor~\cite{Liu:2017man}.
In Eq.~(\ref{eq:CDER-2pt}), $R$ is the 4-dimensional
truncated size of the topological operator, $\delta m$ is the effective mass gap between the nucleon and its excited states, 
and $m_{\eta}$ is the mass of the pseudoscalar meson $\eta$.

Similarly, the three-point function with $Q_t$ can be converted into a four-point function with $q_t(x)$
\begin{align}
G^{(3)Q}
&\sim\sum_{\vec{x}\vec{y}}e^{-i\vec{q}(\vec{x}-\vec{y})}\left\langle\chi\left(x\right)\sum_{r}^{|r|<R}q_{t}\left(y+r\right)J_{\mu}\left(y\right)\bar{\chi}\left(t_{0},{\cal G}\right)\right\rangle\nonumber\\
&\sim M^{(3)Q} + {\cal O}\left(e^{-\delta m(t_c-t_0)},e^{-\delta E(\vec{q})(t_f-t_c)},e^{-m_{\eta}R}\right),\label{eq:CDER-4pt}
\end{align}
where $y=\left(t_c,\vec{y}\right)$, 
and $\delta E(\vec{q})$ is the energy gap of the nucleon and its excited states with 3-momentum $\vec{q}$ at the sink.
Using Eqs.~(\ref{eq:CDER-2pt}) and (\ref{eq:CDER-4pt}), the $\cancel{CP}$ form
factor $F_{3}$ can be calculated as a function of 
cutoff $R$. Due to the cluster decomposition
principle, operators far enough separated have exponentially small correlation. When
the distance between operators is larger than the correlation length $\sim1/m_{\eta}$,
the signal falls below the noise
while the errors still accumulate in the disconnected insertions~\citep{Liu:2017man}. 
So we bind the topological charge to the
sink of the nucleon in the three-point functions or to the inserted currents in the four-point function
to see if a proper cutoff $R$ exists, such that the physics is not
altered while the errors can be reduced. 

\begin{figure}
\begin{centering}
\includegraphics[page=1,width=0.44\textwidth]{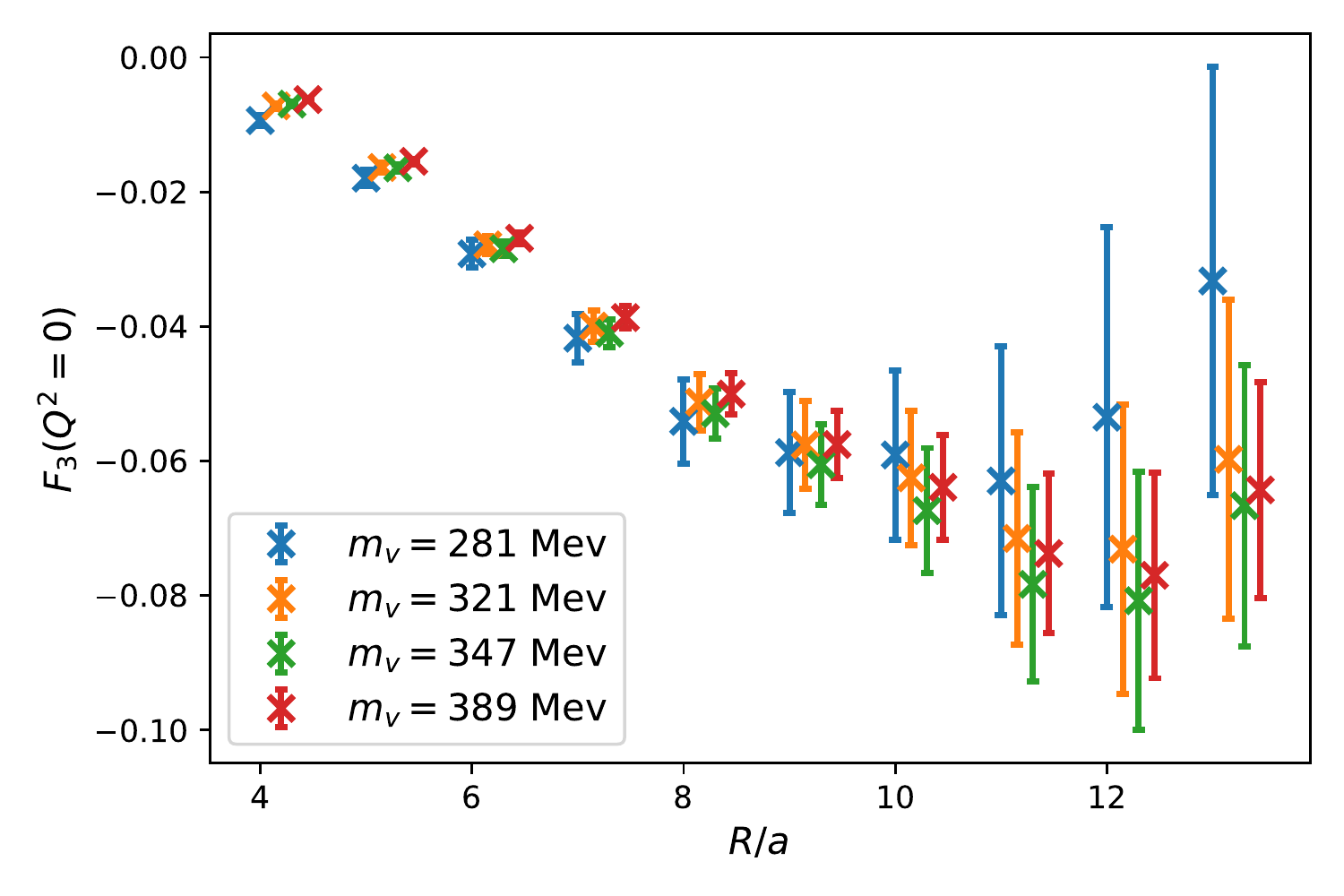}
\par\end{centering}
\centering{}\caption{The cutoff dependence of $F_{3,n}(Q^2=0.2~\mathrm{GeV}^2)$ with different $m_{\pi,v}$ and $m_{\pi,s}=339$~MeV. 
We can see that the value saturates at $R\sim 9a$.
\label{fig:The-cutoff-dependence}}
\end{figure}

Then we do the two-state fit to eliminate the excited-state contamination of nucleon matrix elements at each value of $R$, 
and obtain $F_{3}(Q^2)$ as a function of $R$. 
The corresponding systematic uncertainty
is estimated to be the
difference between the value from the two-state fits and
that from single-exponential fits using only the middle point at different separations.
Taking $F_{3,n}(Q^2=0.2~\mathrm{GeV}^2)$ at $m_{\pi,s}=339$ MeV and different $m_{\pi,v}$ 
as an example 
(shown in Fig.~\ref{fig:The-cutoff-dependence}), 
the central value starts to saturate at around $R=9a\sim 2/m_{\eta}$ as expected.
Since the $R$ dependence for different pion masses are similar, we choose $R_c=9a$
as our optimal cutoff in the neutron case.
For the proton, we use $R_c=10a$.
The systematic uncertainty of this cutoff will be estimated by two independent ways: 
1) taking the difference between the value at the cutoff $R_c$ and the constant fit result with $R\ge R_c$; 
2) fitting the correlation between the topological charge density and the current operator in the nucleon state
to an exponential form first,
and then taking the summation of the correlation in the tail $R\ge R_c$. 
Either way suggests a $\sim$12\% systematic uncertainty. 

\begin{figure}
\begin{centering}
\includegraphics[page=5,width=0.40\textwidth]{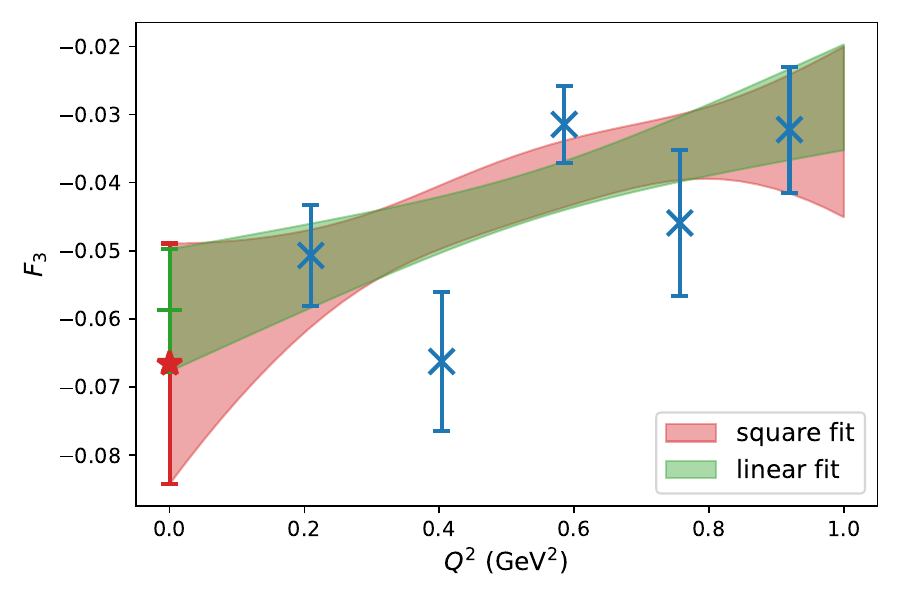}
\par\end{centering}
\centering{}\caption{The $Q^2$ dependence of $F_{3,n}$ with $m_{\pi,v}\sim m_{\pi,s}=339$~MeV.
The green band shows a linear fit in $Q^2$ while the red band shows the fit with an additional $Q^4$ term.
\label{fig:p-extra}
}
\end{figure}

Benefited from CDER, the data points of $F_{3,n}(Q^2)$
show a non-vanishing $Q^2$ dependence 
as shown in Fig~\ref{fig:p-extra} for the case of $m_{\pi,v}\sim m_{\pi,s}=340$~MeV, 
while there is no significant deviation from a linear shape.
Thus we use a linear fit for the extrapolation to $Q^2=0$,
and estimate the corresponding systematic uncertainty
to be the difference between the extrapolated value and
the data value with the smallest $Q^2$.

\begin{figure}
\begin{centering}
\includegraphics[scale=0.48,page=3]{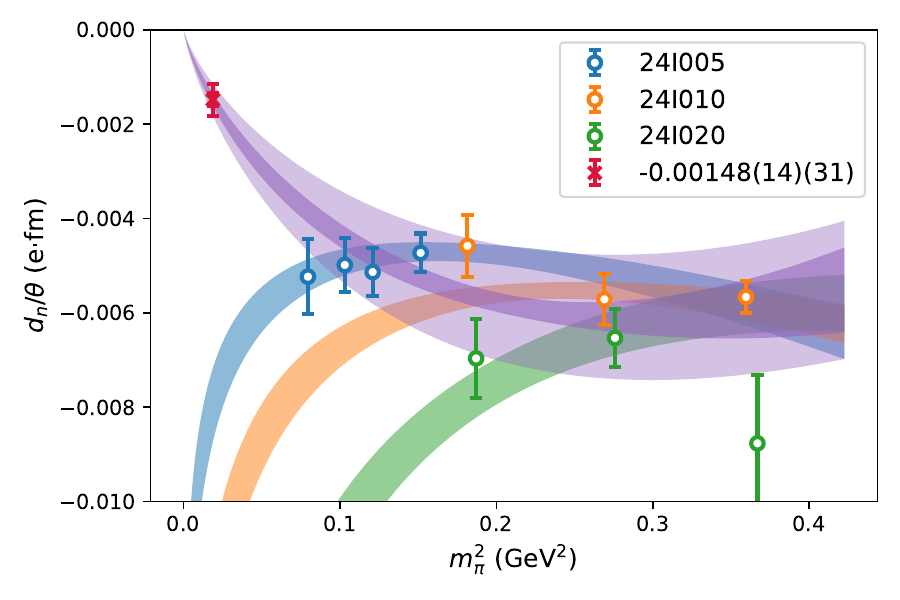}
\par\end{centering}
\begin{centering}
\includegraphics[scale=0.48,page=4]{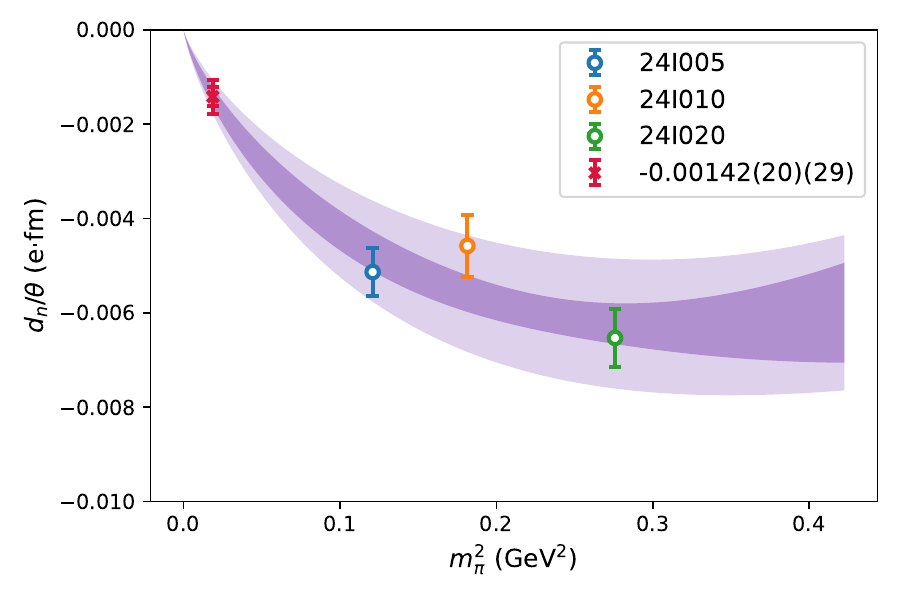}
\par\end{centering}
\centering{}\caption{
The chiral extrapolation of $d_n/{\theta}$ on both the sea and valence quark masses (upper panel) 
and on only the unitary points (lower panel).\label{fig:The-chiral-extrapolation}}
\end{figure}

After the $Q^2\rightarrow0$ extrapolation, 
the final chiral extrapolation of the neutron EDM is shown in the upper panel of Fig.~\ref{fig:The-chiral-extrapolation} with both valence and sea pion mass dependencies. 
We observe that the partially quenched data behave differently from those with unitary points in the lower panel. 
The former tend to move away from zero as the valence quark mass decreases.
Using the overlap fermion allows us to fit our data with the partially quenched chiral perturbation form~\citep{OConnell:2005mfp} 
at finite lattice spacing,
\begin{eqnarray}
d_{n,p} &=& c_{1,n/p}m_{\pi,s}^{2}\log\left(\frac{m_{\pi,v}^{2}}{m_{N}^{2}}\right)+c_{2,n/p}m_{\pi,s}^{2} \nonumber \\
&+& c_{3,n/p}\left(m_{\pi,v}^{2}-m_{\pi,s}^{2}\right),
\end{eqnarray}
where $c_{1,2,3,n/p}$ are free parameters.
Our lattice data are well fitted with $\chi^2/d.o.f.=1.2$, 
and our numerical results suggest that the 
different valence and sea quark mass dependence
is consistent with the chiral perturbation expression.
It is also interesting to point out that the chiral log term is crucial to ensure that the NEDM approaches zero 
in the chiral limit of both the valence and sea quark masses.
With the zero NEDM constraint at the chiral limit,
our interpolated result for neutron is $d_{n}=-0.00148(14)$,
where the statistical uncertainty is less than 10\%.
This is quite an improvement from the 2 $\sigma$ statistical error in Ref.~\cite{Dragos:2019oxn}.

We also carry out another chiral extrapolation using only the unitary pion mass points,
as shown in the lower panel of Fig.~\ref{fig:The-chiral-extrapolation}.
It gives $d_{n}=-0.00142(20) \bar{\theta}$,
which is consistent with the prediction using partially quenched data points 
but with larger statistical uncertainty.
We take the difference between the extrapolated results with and without partially quenched data
points as an estimation of the systematic uncertainty in the chiral extrapolation. 

The proton EDM and its systematic uncertainties can be obtained with a similar procedure. 
More detailed discussion on the fits, systematic uncertainty estimation, and 
proton EDM can be found in the Supplemental Materials~\cite{supplemental}. 

\textit{Summary:} 
We calculate the nucleon electric dipole moment with overlap fermions
on 3 domain wall lattices at lattice spacing 0.11 fm. Since the
overlap fermion preserves chiral symmetry, 
we have well-defined topological charge and
the chiral extrapolation
is carried out reliably without the need of doing continuum extrapolations
first. 
We have in total 3 sea pion masses and 10 partially quenched
valence pion masses in the chiral fitting and find that the EDM dependence
on the sea and valence pion masses behaves oppositely.

With the help of the cluster decomposition error reduction (CDER) technique,
we determine the neutron and proton EDM at the physical pion mass point
to be 
$d_{n}=-0.00148\left(14\right)\left(31\right)\bar\theta$ e$\cdot$fm
and 
$d_{p}=0.0038\left(11\right)\left(8\right)\bar\theta$ e$\cdot$fm,
respectively.
The two uncertainties are the statistical uncertainty and the total systematic uncertainty
from the excited-state contamination, the CDER cutoff, and the $Q^2$ and chiral extrapolations.
By using the most recent experimental upper limit of $d_{n}$, our
results indicate that $\bar{\theta}<10^{-10}$. 
This work demonstrates
the advantage 
of using chiral fermions 
in the NEDM calculation
and paves the road to future precise studies of the strong $\cancel{CP}$
effects.

\begin{acknowledgments}
\section*{Acknowledgments}
JL is supported by Guangdong Major Project of Basic and Applied Basic Research under Grant No.\ 2020B0301030008, 
Science and Technology Program of Guangzhou under Grant No.\ 2019050001, 
and the Natural Science Foundation of China (NSFC) under Grant No.\ 12175073 and No.\ 12222503.
TD and KL are supported in part by the Office of Science of the U.S.\ Department of Energy under 
Grant No.\ DE-SC0013065 (TD and KL) and No.\ DE-AC05-06OR23177 (KL), which is within the framework of the TMD Topical Collaboration.
YY is supported in part by the Strategic Priority Research Program of Chinese Academy of Sciences, 
Grant No.\ XDB34030303 and XDPB15, NSFC under Grant No.\ 12293062, and also a NSFC-DFG joint grant under Grant No.\ 12061131006 and SCHA~458/22.
GW is supported by the French National Research Agency under the contract ANR-20-CE31-0016.
AA is supported in part by U.S.\ DOE Grant No.\ DE-FG02-95ER40907.
This research used resources of the Oak Ridge Leadership Computing Facility at the Oak Ridge National Laboratory, 
which is supported by the Office of Science of the U.S.\ Department of Energy under Contract No.\ DE-AC05-00OR22725. 
This work used Stampede time under the Extreme Science and Engineering Discovery Environment (XSEDE), 
which is supported by National Science Foundation Grant No.\ ACI-1053575. 
We also used resources on Frontera at Texas Advanced Computing Center (TACC). 
The analysis work is partially done on the supercomputing system in the Southern Nuclear Science Computing Center (SNSC).
We also thank the National Energy Research Scientific Computing Center (NERSC) for providing HPC resources that have contributed to the research results reported within this paper.
We acknowledge the facilities of the USQCD Collaboration used for this research in part, which are funded by the Office of Science of the U.S.\ Department of Energy.
\end{acknowledgments}

\bibliographystyle{unsrt}
\bibliography{library}

\begin{thebibliography}{10}

\bibitem{Sakharov:1967dj}
A.D. Sakharov.
\newblock {Violation of CP Invariance, C asymmetry, and baryon asymmetry of the
  universe}.
\newblock {\em Sov. Phys. Usp.}, 34(5):392--393, 1991.

\bibitem{Farrar:1993sp}
Glennys~R. Farrar and M.E. Shaposhnikov.
\newblock {Baryon asymmetry of the universe in the minimal Standard Model}.
\newblock {\em Phys. Rev. Lett.}, 70:2833--2836, 1993.
\newblock [Erratum: Phys.Rev.Lett. 71, 210 (1993)].

\bibitem{Farrar:1993hn}
Glennys~R. Farrar and M.E. Shaposhnikov.
\newblock {Baryon asymmetry of the universe in the standard electroweak
  theory}.
\newblock {\em Phys. Rev. D}, 50:774, 1994.

\bibitem{Gavela:1993ts}
M.B. Gavela, P.~Hernandez, J.~Orloff, and O.~Pene.
\newblock {Standard model CP violation and baryon asymmetry}.
\newblock {\em Mod. Phys. Lett. A}, 9:795--810, 1994.

\bibitem{Gavela:1994dt}
M.B. Gavela, P.~Hernandez, J.~Orloff, O.~Pene, and C.~Quimbay.
\newblock {Standard model CP violation and baryon asymmetry. Part 2: Finite
  temperature}.
\newblock {\em Nucl. Phys. B}, 430:382--426, 1994.

\bibitem{Huet:1994jb}
Patrick Huet and Eric Sather.
\newblock {Electroweak baryogenesis and standard model CP violation}.
\newblock {\em Phys. Rev. D}, 51:379--394, 1995.

\bibitem{Smith:1957ht}
J.H. Smith, E.M. Purcell, and N.F. Ramsey.
\newblock {Experimental limit to the electric dipole moment of the neutron}.
\newblock {\em Phys. Rev.}, 108:120--122, 1957.

\bibitem{Abel:2020gbr}
C.~Abel et~al.
\newblock {Measurement of the permanent electric dipole moment of the neutron}.
\newblock {\em Phys. Rev. Lett.}, 124(8):081803, 2020.

\bibitem{Abramczyk:2017oxr}
M.~Abramczyk, S.~Aoki, T.~Blum, T.~Izubuchi, H.~Ohki, and S.~Syritsyn.
\newblock {Lattice calculation of electric dipole moments and form factors of
  the nucleon}.
\newblock {\em Phys. Rev. D}, 96(1):014501, 2017.

\bibitem{Shintani:2005xg}
E.~Shintani, S.~Aoki, N.~Ishizuka, K.~Kanaya, Y.~Kikukawa, Y.~Kuramashi,
  M.~Okawa, Y.~Tanigchi, A.~Ukawa, and T.~Yoshie.
\newblock {Neutron electric dipole moment from lattice QCD}.
\newblock {\em Phys. Rev.}, D72:014504, 2005.

\bibitem{Berruto:2005hg}
F.~Berruto, T.~Blum, K.~Orginos, and A.~Soni.
\newblock {Calculation of the neutron electric dipole moment with two dynamical
  flavors of domain wall fermions}.
\newblock {\em Phys. Rev.}, D73:054509, 2006.

\bibitem{Guo:2015tla}
F.~K. Guo, R.~Horsley, U.~G. Meissner, Y.~Nakamura, H.~Perlt, P.~E.~L. Rakow,
  G.~Schierholz, A.~Schiller, and J.~M. Zanotti.
\newblock {The electric dipole moment of the neutron from 2+1 flavor lattice
  QCD}.
\newblock {\em Phys. Rev. Lett.}, 115(6):062001, 2015.

\bibitem{Shintani:2015vsx}
Eigo Shintani, Thomas Blum, Taku Izubuchi, and Amarjit Soni.
\newblock {Neutron and proton electric dipole moments from $N_f=2+1$
  domain-wall fermion lattice QCD}.
\newblock {\em Phys. Rev.}, D93(9):094503, 2016.

\bibitem{Alexandrou:2015spa}
C.~Alexandrou, A.~Athenodorou, M.~Constantinou, K.~Hadjiyiannakou, K.~Jansen,
  G.~Koutsou, K.~Ottnad, and M.~Petschlies.
\newblock {Neutron electric dipole moment using $N_{f} =2+1+1$ twisted mass
  fermions}.
\newblock {\em Phys. Rev.}, D93(7):074503, 2016.

\bibitem{Syritsyn:2019vvt}
Sergey Syritsyn, Taku Izubuchi, and Hiroshi Ohki.
\newblock {Calculation of Nucleon Electric Dipole Moments Induced by Quark
  Chromo-Electric Dipole Moments and the QCD $\theta$-term}.
\newblock {\em PoS}, Confinement2018:194, 2019.

\bibitem{Dragos:2019oxn}
Jack Dragos, Thomas Luu, Andrea Shindler, Jordy de~Vries, and Ahmed Yousif.
\newblock {Confirming the Existence of the strong CP Problem in Lattice QCD
  with the Gradient Flow}.
\newblock {\em Phys. Rev. C}, 103(1):015202, 2021.

\bibitem{Alexandrou:2020mds}
C.~Alexandrou, A.~Athenodorou, K.~Hadjiyiannakou, and A.~Todaro.
\newblock {Neutron electric dipole moment using lattice QCD simulations at the
  physical point}.
\newblock {\em Phys. Rev. D}, 103(5):054501, 2021.

\bibitem{Bhattacharya:2021lol}
Tanmoy Bhattacharya, Vincenzo Cirigliano, Rajan Gupta, Emanuele Mereghetti, and
  Boram Yoon.
\newblock {Contribution of the QCD $\Theta$-term to the nucleon electric dipole
  moment}.
\newblock {\em Phys. Rev. D}, 103(11):114507, 2021.

\bibitem{supplemental}
{\em Supplemental Material}.

\bibitem{Baluni:1978rf}
Varouzhan Baluni.
\newblock {CP Violating Effects in QCD}.
\newblock {\em Phys. Rev. D}, 19:2227--2230, 1979.

\bibitem{Aoki:2010dy}
Y.~Aoki et~al.
\newblock {Continuum Limit Physics from 2+1 Flavor Domain Wall QCD}.
\newblock {\em Phys.Rev.}, D83:074508, 2011.

\bibitem{Neuberger:1997fp}
Herbert Neuberger.
\newblock {Exactly massless quarks on the lattice}.
\newblock {\em Phys. Lett.}, B417:141--144, 1998.

\bibitem{Hasenfratz:2002rp}
P.~Hasenfratz, S.~Hauswirth, T.~Jorg, F.~Niedermayer, and K.~Holland.
\newblock {Testing the fixed point QCD action and the construction of chiral
  currents}.
\newblock {\em Nucl. Phys.}, B643:280--320, 2002.

\bibitem{Li:2010pw}
A.~Li et~al.
\newblock {Overlap Valence on 2+1 Flavor Domain Wall Fermion Configurations
  with Deflation and Low-mode Substitution}.
\newblock {\em Phys. Rev. D}, 82:114501, 2010.

\bibitem{Yang:2015zja}
Yi-Bo Yang, Andrei Alexandru, Terrence Draper, Ming Gong, and Keh-Fei Liu.
\newblock {Stochastic method with low mode substitution for nucleon isovector
  matrix elements}.
\newblock {\em Phys. Rev.}, D93(3):034503, 2016.

\bibitem{Liu:2017man}
Keh-Fei Liu, Jian Liang, and Yi-Bo Yang.
\newblock {Variance Reduction and Cluster Decomposition}.
\newblock {\em Phys. Rev.}, D97(3):034507, 2018.

\bibitem{Adams:1998eg}
David~H. Adams.
\newblock {Axial anomaly and topological charge in lattice gauge theory with
  overlap Dirac operator}.
\newblock {\em Annals Phys.}, 296:131--151, 2002.

\bibitem{Fujikawa:1998if}
Kazuo Fujikawa.
\newblock {A Continuum limit of the chiral Jacobian in lattice gauge theory}.
\newblock {\em Nucl. Phys. B}, 546:480--494, 1999.

\bibitem{OConnell:2005mfp}
Donal O'Connell and Martin~J. Savage.
\newblock {Extrapolation formulas for neutron EDM calculations in lattice QCD}.
\newblock {\em Phys. Lett. B}, 633:319--324, 2006.

\bibitem{Note1}
One way to understand this is to think of the nucleon two-point function. The
  term $i\theta {\protect \rm Tr}\left [\protect \frac {1+\gamma _4}{2}\langle
  \chi Q_t \protect \bar \chi \rangle \right ]$ vanishes due to the $0^{-+}$
  quantum number of $Q_t$.

\bibitem{Liu:2002qu}
Keh-Fei Liu.
\newblock {Heavy and light quarks with lattice chiral fermions}.
\newblock {\em Int. J. Mod. Phys.}, A20:7241--7254, 2005.

\bibitem{Liang:2018pis}
Jian Liang, Yi-Bo Yang, Terrence Draper, Ming Gong, and Keh-Fei Liu.
\newblock {Quark spins and Anomalous Ward Identity}.
\newblock {\em Phys. Rev.}, D98(7):074505, 2018.

\bibitem{Muller-Preussker:2015daa}
M.~M{\"u}ller-Preussker.
\newblock {Recent results on topology on the lattice (in memory of Pierre van
  Baal)}.
\newblock {\em PoS}, LATTICE2014:003, 2015.

\bibitem{Luscher:2010iy}
Martin L{\"u}scher.
\newblock {Properties and uses of the Wilson flow in lattice QCD}.
\newblock {\em JHEP}, 1008:071, 2010.

\bibitem{Luscher:2011bx}
Martin Luscher and Peter Weisz.
\newblock {Perturbative analysis of the gradient flow in non-abelian gauge
  theories}.
\newblock {\em JHEP}, 1102:051, 2011.

\bibitem{Luscher:2013cpa}
Martin Luscher.
\newblock {Chiral symmetry and the Yang--Mills gradient flow}.
\newblock {\em JHEP}, 1304:123, 2013.

\end{thebibliography}

\begin{widetext}






\begin{center}
{\LARGE\bf{Supplemental Materials}}\par
\end{center}

\section{Conventions and Formalism\label{sec:Conventions}}

In this part of Supplemental Materials, we list our notations
and conventions in a very detailed manner, which we think is
quite worthwhile since the final sign of
EDM depends directly on the conventions used. 

\subsection{Gamma Matrices}

First, for the gamma matrices in Minkowski space, we use

\begin{equation}
\left\{\gamma_{\mu},\gamma_{\nu}\right\} =2\eta_{\mu\nu},
\end{equation}
where $\eta_{\mu\nu}=\left(+,-,-,-\right)$ is the corresponding metric tensor. Similarly, we have, for the Euclidean
ones,
\begin{equation}
\left\{ \gamma_{\mu}^{E},\gamma_{\nu}^{E}\right\} =2\eta_{\mu\nu}^{E},
\end{equation}
with $\eta_{\mu\nu}^{E}=\left(+,+,+,+\right)$.
Our choice is to let $\gamma_{4}^{E}=\gamma^{0}$ while $\gamma_{i}^{E}=-{i}\gamma^{i}$.
For the momentum we have $p_{4}^{E}={i}E={i}p^{0}$ and $p_{i}^{E}=p^{i}$,
this definition ensures $p\!\!\!/=\gamma_{0}p^{0}+\gamma_{i}p^{i}=-{i}\gamma_{4}^{E}p_{4}^{E}-{i}\gamma_{i}^{E}p_{i}^{E}=-{i}p\!\!\!/^{E}$. 

Then, with the above definitions, we come to the following convention
of the spinors
\begin{equation}
u\bar{u}=\frac{p\!\!\!/+m}{2m},\ u^{E}\bar{u}^{E}=\frac{-ip\!\!\!/^{E}+m}{2m},
\end{equation}
and we define 
\begin{equation}
\sigma_{\mu\nu}=\frac{{i}}{2}\left[\gamma_{\mu},\gamma_{\nu}\right], \sigma_{\mu\nu}^{E}=\frac{1}{2i}\left[\gamma_{\mu}^{E},\gamma_{\nu}^{E}\right].
\end{equation}
in our notations.

\subsection{QCD Lagrangian with the $\theta$ Term}

The Minkowski QCD Lagrangian reads 
\begin{equation}
{\cal L}=\bar{\psi}\left({i}D\!\!\!\!/-m\right)\psi-\frac{1}{4}F_{\mu\nu}^{a}F_{a}^{\mu\nu}=\bar{\psi}\left({i}D\!\!\!\!/-m\right)\psi-\frac{1}{2}\mathrm{Tr}[F_{\mu\nu}F^{\mu\nu}],
\end{equation}
where the covariant derivative is $D_{\mu}\equiv\partial_{\mu}-{i}gA_{\mu}$
with a minus sign in front of $A_{\mu}$.
Along with this convention,
we use
\begin{equation}
F_{\mu\nu}\equiv \frac{1}{-{i}g}[D_{\mu},D_{\nu}]=\partial_{\mu}A_{\nu}-\partial_{\nu}A_{\mu}-{i}g[A_{\mu},A_{\nu}].
\end{equation}

To have the QCD Lagrangian in Euclidean space, we first notice $\partial^{0}=i\partial_{4}^{E}$ and $\partial^{i}=-\partial_{i}^{E}$.
And for the gauge fields, the conversion is the same as that of $p^E$ and $p$:
\begin{equation}
A_{4}^{E}={i}A^{0},\ A_{i}^{E}=A^{i}.
\end{equation}
Combining the above relations, we come to
\begin{equation}
D^{0}=\frac{\partial}{\partial x_{0}}-{i}gA^{0}\to {i}\left(\frac{\partial}{\partial x_{4}^{E}}+{i}gA_{4}^{E}\right)\equiv {i}D_{4}^{E},
\end{equation}
and
\begin{equation}
D^{i}=\frac{\partial}{\partial x_{i}}-{i}gA^{i}\to-\left(\frac{\partial}{\partial x_{i}^{E}}+{i}gA_{i}^{E}\right)\equiv-D_{i}^{E}.
\end{equation}
Plugging in the conversions of the gamma matrices, we have 
\begin{equation}
{i}D^{0}\gamma_{0}+{i}D^{i}\gamma_{i}-m=-D_{4}^{E}\gamma_{4}^{E}-D_{i}^{E}\gamma_{i}^{E}-m.
\end{equation}
The Minkowski field tensor satisfies
\begin{equation}
F_{\mu\nu}F^{\mu\nu}=2\sum F_{0i}F^{0i}+2\sum_{i<j}F_{ij}F^{ij}=-2E^{2}+2B^{2},
\end{equation}
where $E^{i}=E_{i}=F_{0i}=-F^{0i}$, and $B^{i}=-\frac{1}{2}\epsilon^{ijk}F_{jk}=B_{i}=-\frac{1}{2}\epsilon_{ijk}F^{jk}$.
It is easy to check that
\begin{equation}
E^{i}=-{i}E_{i}^{E}, B^{i}=-B_{i}^{E},
\end{equation}
and such that
\begin{equation}
F_{\mu\nu}^{E}F^{E,\mu\nu}=2\left[\left(E^{E}\right)^{2}+\left(B^{E}\right)^{2}\right]=F_{\mu\nu}F^{\mu\nu}.
\end{equation}
Then, we finally reach the form of the QCD Lagrangian in Euclidean
space
\begin{equation}
{\cal L}^{E}=\bar{\psi}\left(D\!\!\!\!/^{E}+m\right)\psi+\frac{1}{2}\mathrm{Tr}\left[F_{\mu\nu}^{E}F^{E,\mu\nu}\right].
\end{equation}

When the $\theta$ term is taken into consideration, in Minkowski
space, we have ${\cal L}\to{\cal L}+\mathcal{L}_{\theta}$ and
\begin{equation}
\mathcal{L}_{\theta}=\bar{\theta}\frac{g^{2}}{32\pi^{2}}F_{\mu\nu}^{a}\tilde{F}_{a}^{\mu\nu}=\bar{\theta}\frac{g^{2}}{16\pi^{2}}\mathrm{Tr}\left[F_{\mu\nu}\tilde{F}^{\mu\nu}\right]\equiv\bar{\theta}q_{t}
\end{equation}
where $\tilde{F}^{\mu\nu}=\epsilon^{\mu\nu\rho\sigma}F_{\rho\sigma}$
and $q_{t}$ is the topological charge density. Based on the above conversions, we have
\begin{align}
F_{\mu\nu}\tilde{F}^{\mu\nu}&=2\sum F_{0i}\tilde{F}^{0i}+2\sum_{i<j}F_{ij}\tilde{F}^{ij}=-8E\cdot B,\nonumber\\
F_{\mu\nu}^{E}\tilde{F}^{E,\mu\nu}&=2\sum F_{0i}^{E}\tilde{F}^{E,0i}+2\sum_{i<j}F_{ij}^{E}\tilde{F}^{E,ij}=8{i}E\cdot B,
\end{align}
and in the end
\begin{equation}
{\cal L}^{E}+\mathcal{L}_{\theta}^E=\bar{\psi}\left(D\!\!\!\!/^{E}+m\right)\psi+
\frac{1}{2}\mathrm{Tr}\left[F_{\mu\nu}^{E}F^{E,\mu\nu}\right]-{i}\bar{\theta}\frac{g^{2}}{16\pi^{2}}\mathrm{Tr}\left[F_{\mu\nu}^{E}\tilde{F}^{E,\mu\nu}\right].
\end{equation}

\subsection{Spinors Under the $\theta$ Vacuum}
Now we have determined the Lagrangian in Euclidean space. In the following
part of the Supplemental Materials, we will work in the Euclidean
space and omit the superscript $^E$ unless otherwise specified.

After the $\theta$ term is plugged in, the $P$ and $CP$ symmetries
are broken. The normal Dirac equation
and spinor definition
should be modified. The new Dirac equation reads 
\begin{equation}
\left[-{i}p\!\!\!/-m^{\theta}e^{-{i}\alpha\left(\theta\right)\gamma_{5}}\right]u^{\theta}=
\bar{u}^{\theta}\left[-{i}p\!\!\!/-m^{\theta}e^{-{i}\alpha\left(\theta\right)\gamma_{5}}\right]=0,
\end{equation}
where the superscript $\theta$ denotes quantities under the $\theta$
vacuum and $\alpha\left(\theta\right)$ is an unknown function of $\theta$.
Up to terms linear in $\theta$ (due to the smallness of $\theta$),
we have, for example,
\begin{equation}
\left[-{i}p\!\!\!/-m\left(1+f_{m}^{1}\theta\right)\left(1-{i}\alpha^{1}\theta\gamma_{5}\right)\right]\left(1+f_{u}^{1}\theta\right)u=0,
\end{equation}
where $f_{m}^{1}$, $\alpha^{1}$, and $f_{u}^{1}$ are expansion coefficients.
Subtracting the normal Dirac equation, we get 
\begin{equation}
-m\left(f_{m}^{1}-{i}\alpha^{1}\gamma_{5}\right)u+\left[-{i}p\!\!\!/-m\right]f_{u}^{1}u=0.
\end{equation}
Since the nucleon mass has no leading $\theta$ correction\footnote{
One way to understand this is to think of the nucleon two-point function. The term
$i\theta{\rm Tr}\left[\frac{1+\gamma_4}{2}\langle \chi  Q_t \bar\chi \rangle\right]$ vanishes
due to the $0^{-+}$ quantum number of $Q_t$.}
\begin{equation}
m^{\theta}=m+{\cal O}\left(\theta^{2}\right),
\end{equation}
the new spinors can be expressed as
\begin{equation}
u^{\theta}=e^{i\alpha^{1}\theta\gamma_{5}}u,
\end{equation}
and
\begin{equation}
\bar{u}^{\theta}=\bar{u}e^{i\alpha^{1}\theta\gamma_{5}},
\end{equation}
such that we have
\begin{equation}
u^{\theta}(p)\bar{u}^{\theta}(p)=\frac{-ip\!\!\!/+me^{i2\alpha^{1}\gamma_{5}\theta}}{2m}.
\end{equation}
Also, we define the overlapping factor
\begin{equation}
\left\langle 0|\chi|N\right\rangle =Zu,
\end{equation}
where $\chi$ is the nucleon interpolating filed operator and $\left|N\right\rangle $
is the corresponding nucleon state. Then, under the $\theta$ vacuum
we define
\begin{equation}
_{\theta}\langle0|\chi|N\rangle_{\theta}=Z^{\theta}u^{\theta}.
\end{equation}
Similarly, we have
\begin{equation}
Z^{\theta}=Z+{\cal O}\left(\theta^{2}\right).
\end{equation}

\subsection{Form Factors\label{sec:form-factors}}
In Minkowski space, we use the following electromagnetic form factor decomposition
\begin{equation}
\left\langle N'|\bar\psi\gamma_{\mu}\psi|N\right\rangle =\bar{u}\left(p'\right)\left[\gamma_{\mu}F_{1}(q^{2})+i\sigma_{\mu\nu}q^{\nu}\frac{F_{2}(q^{2})}{2m}\right]u\left(p\right),
\end{equation}
where $F_{1}$ and $F_{2}$ are the Pauli and Dirac form factors
respectively, $q=p'-p$ with $p'$ the momentum of the outgoing nucleon
($\bar{u}\left(p'\right)$) and $p$ the momentum of the incoming
nucleon. For the Minkowski case, with our conventions we have
\begin{equation}
i\sigma_{\mu\nu}q^{\nu}=\left(q_{\mu}-\gamma_{\mu}q\!\!\!/\right),
\end{equation}
and using the Dirac equation $\left(p\!\!\!/-m\right)u=0$ we get
\begin{align}
 & \bar{u}\left(p'\right)\left[i\sigma_{\mu\nu}q^{\nu}\right]u\left(p\right)\nonumber \\
= & 2m\bar{u}\left(p'\right)\left[\gamma_{\mu}\right]u\left(p\right)-\bar{u}\left(p'\right)\left[p'_{\mu}+p_{\mu}\right]u\left(p\right).
\end{align}
On the other hand, with the Euclidean notation, we have $-\sigma_{\mu\nu}^{E}q_{\nu}^{E}=i\left(\gamma_{\mu}^{E}q\!\!\!/^{E}-q_{\mu}^{E}\right)$.
And similarly
\begin{align}
 & \bar{u}^{E}\left(p'^{E}\right)\left[-\sigma_{\mu\nu}^{E}q_{\nu}^{E}\right]u^{E}\left(p^{E}\right)\nonumber \\
= & 2m\bar{u}^{E}\left(p'^{E}\right)\left[\gamma_{\mu}^{E}\right]u^{E}\left(p\right)+i\bar{u}^{E}\left(p'\right)\left[p_{\mu}'^{E}+p_{\mu}^{E}\right]u^{E}\left(p^{E}\right).
\end{align}
So in order to have consistent results for both Minkowski and Euclidean
space, one should use $-\sigma_{\mu\nu}^{E}q_{\nu}^{E}$ under our
convention:
\begin{equation}
\left\langle N'|\gamma_{\mu}|N\right\rangle ^{E}=\bar{u}^{E}\left(p'^{E}\right)\left[\gamma_{\mu}^{E}F_{1}(q^{2})-\sigma_{\mu\nu}^{E}q_{\nu}^{E}\frac{F_{2}(q^{2})}{2m}\right]u^{E}\left(p^{E}\right).
\end{equation}
For the $\cancel{CP}$ case, we have an additional form factor $F'_3$
\begin{equation}
-\sigma_{\mu\nu}q_{\nu}\gamma_{5}\frac{F'_{3}(q^{2})}{2m}.
\end{equation}
{\it{N.B.}}, when taking the phase carried by
the $\cancel{CP}$ spinors into consideration, this $CP$ odd form
factor should be modified as well.
The relation
between the correct $\cancel{CP}$ form factor under the $\theta$ vacuum $F_{3}$ 
and $F_{3}'$ can be retrieved by considering the
parity transformation of the normal spinors 
\begin{equation}
u\left(p\right)\to u\left(\tilde{p}\right)=\gamma_{4}u\left(p\right),\bar{u}\left(p\right)\to\bar{u}\left(\tilde{p}\right)=\bar{u}\left(p\right)\gamma_{4},
\end{equation}
and the $\cancel{CP}$ ones

\begin{equation}
u^{\theta}\left(p\right)\to u^{\theta}\left(\tilde{p}\right)=e^{i\alpha_{1}\theta\gamma_{5}}\gamma_{4}u\left(p\right)=\left(1+i\alpha_{1}\theta\gamma_{5}\right)\gamma_{4}u,
\end{equation}
\begin{equation}
\bar{u}^{\theta}\left(p\right)\to\bar{u}^{\theta}\left(\tilde{p}\right)=\bar{u}\left(p\right)\gamma_{4}e^{i\alpha^{1}\theta\gamma_{5}}=\bar{u}\left(p\right)\gamma_{4}\left(1+i\alpha_{1}\theta\gamma_{5}\right).
\end{equation}
Specifically, we have 
\begin{equation}
i\theta F_{3}=2i\alpha^{1}\theta F_{2}+i\theta F_{3}'=i\theta\left(2\alpha^{1}F_{2}+F_{3}'\right).
\end{equation}

\subsection{Correlation Functions}
In general, path integrals under the $\theta$ vacuum can be estimated
by employing the Taylor expansion in $\theta$ and keeping only the
\purple{leading term}
\begin{align}
 & \int DA\cdot{\rm Det}\left[M\right]e^{-S_{g}+i\theta Q_{t}}\nonumber \\
\sim & \int DA\cdot{\rm Det}\left[M\right]e^{-S_{g}}+i\theta\int DA\cdot{\rm Det}\left[M\right]Q_{t}e^{-S_{g}},
\end{align}
where $Q_{t}=\int d^{4}xq_{t}=\frac{g^{2}}{32\pi^{2}}\int d^{4}xF_{\mu\nu}^{E}\tilde{F}^{E,\mu\nu}$
is the total topological charge and $q_{t}$ is the charge density.
Correlation functions can therefore be accessed by
\begin{equation}
_{\theta}\langle...\rangle_{\theta}=\langle...\rangle+i\theta\langle...Q_{t}\rangle.
\end{equation}
For example, the two-point functions can be expressed as
\begin{equation}
G_{2}^{\theta}=G_{2}+i\theta G_{2}^{Q},
\end{equation}
where $G_{2}^{\theta}$, $G_{2}$, and $G_{2}^{Q}$ are two-point
functions evaluated with the $\theta$ term, normal two-point functions,
and two-point functions weighted by the topological charge, respectively. Since, 
\begin{align}
G_{2}^{\theta} & =ZZ'^{\dagger}e^{-Et}\frac{m}{E}u^{\theta}(p)\bar{u}^{\theta}(p)\nonumber \\
 & =ZZ'^{\dagger}e^{-Et}\frac{m}{E}u\bar{u}+ZZ'^{\dagger}e^{-Et}\frac{m}{E}i\alpha^{1}\gamma_{5}\theta,
\end{align}
where $Z$ and $Z'$ are the sink and source overlapping factors and
$m$ and $E$ are the nucleon mass and energy, and
\begin{equation}
G_{2}=ZZ'^{\dagger}e^{-Et}\frac{m}{E}u(p)\bar{u}(p),
\end{equation}
we can get
\begin{equation}
G_{2}^{Q}=G_{2}^{\theta}-G_{2}=ZZ'^{\dagger}e^{-Et}\frac{m}{E}\alpha^{1}\gamma_{5}.
\end{equation}
Here we are assuming $t$ is large enough so that only the ground state
survives to simplify the equations. 
These two-point correlation functions
offer to a way of determining the $\cancel{CP}$ angle $\alpha^{1}$:
\begin{equation}
\frac{1}{2}\frac{{\rm Tr}\left[\gamma_{5}G_{2}^{Q}\right]}{{\rm Tr}\left[\Gamma_{e}G_{2}\right]}=\frac{1}{2}\frac{\alpha^{1}{\rm Tr}\left[I_{4}\right]}{\frac{1}{2}{\rm Tr}\left[I_{4}\right]}=\alpha^{1},\label{eq:alpha}
\end{equation}
where $\Gamma_{e}=\frac{1+\gamma_{4}}{2}$ is the unpolarized projector
and $I_{4}$ is the 4 by 4 identity matrix. The angle $\alpha^{1}$
is actually the leading coefficient of the spinor dependence on $\theta$,
which, in some sense, measures the $\cancel{CP}$ effect of the $\theta$
term.

For the three-point function case, similarly, we have

\begin{equation}
G_{3}^{\theta}=G_{3}+i\theta G_{3}^{Q}.
\end{equation}
The normal three-point function is
\begin{equation}
G_{3}=ZZ'^{\dagger}e^{-E_{f}(t_f-t_c)}e^{-E_{i}t_c}\frac{m^{2}}{E_{f}E_{i}}u(p_{f})\langle N_{f}|J_{\mu}|N_{i}\rangle\bar{u}(p_{i}),
\end{equation}
where the subscripts $i$ and $f$ are for the initial and final nucleons
respectively. Denoting the common factor $ZZ'^{\dagger}e^{-E_{f}(t_f-t_c)}e^{-E_{i}t_c}\frac{m^{2}}{E_{f}E_{i}}=A$
for simplicity, we have
\begin{equation}
G_{3}^{\theta}=Au^{\theta}(p_{f}){}_{\theta}\langle N_{f}|J_{\mu}|N_{i}\rangle_{\theta}\bar{u}^{\theta}(p_{i}).
\end{equation}
The relation between the
correlators and the form factors will be derived as follows. In general,
the nucleon matrix elements in the three-point correlation functions
can be decomposed into $CP$ even and $CP$ odd form factors $W_{\mu}^{{\rm even}}$
and $W_{\mu}^{{\rm odd}}$ as
\begin{equation}
G_{3}=Au(p_{f})\bar{u}(p_{f})W_{\mu}^{{\rm even}}u(p_{i})\bar{u}(p_{i}),
\end{equation}
and
\begin{equation}
G_{3}^{\theta}=Au^{\theta}(p_{f})\bar{u}^{\theta}(p_{f})\left(W_{\mu}^{{\rm even}}+i\theta W_{\mu}^{{\rm odd}}\right)u^{\theta}(p_{i})\bar{u}^{\theta}(p_{i}).
\end{equation}
Thus we have
\begin{equation}
\begin{split}\frac{G_{3}}{A}= & \left(\frac{-ip\!\!\!/_{f}+m}{2m}W_{\mu}^{{\rm even}}\frac{-ip\!\!\!/_{i}+m}{2m}\right)\end{split},
\end{equation}
and 
\begin{equation}
\begin{split}\frac{G_{3}^{\theta}}{A} & =\frac{-ip\!\!\!/_{f}+me^{i2\alpha^{1}\theta\gamma_{5}}}{2m}\left(W_{\mu}^{{\rm even}}+i\theta W_{\mu}^{{\rm odd}}\right)\frac{-ip\!\!\!/_{i}+me^{i2\alpha^{1}\theta\gamma_{5}}}{2m}\\
 & =\left(\frac{-ip\!\!\!/_{f}+m}{2m}W_{\mu}^{{\rm even}}\frac{-ip\!\!\!/_{i}+m}{2m}\right)\\
 & +i\theta \left(\alpha^{1}\gamma_{5}W_{\mu}^{{\rm even}}\frac{-ip\!\!\!/_{i}+m}{2m}+\frac{-ip\!\!\!/_{f}+m}{2m}W_{\mu}^{{\rm even}}\alpha^{1}\gamma_{5}+\frac{-ip\!\!\!/_{f}+m}{2m}W_{\mu}^{{\rm odd}}\frac{-ip\!\!\!/_{i}+m}{2m}\right).
\end{split}
\end{equation}
So by doing a similar subtraction, we arrive at 
\begin{equation}
\frac{G_{3}^{Q}}{A}=\alpha^{1}\gamma_{5}W_{\mu}^{{\rm even}}\frac{-ip\!\!\!/_{i}+m}{2m}+\frac{-ip\!\!\!/_{f}+m}{2m}W_{\mu}^{{\rm even}}\alpha^{1}\gamma_{5}+\frac{-ip\!\!\!/_{f}+m}{2m}W_{\mu}^{{\rm odd}}\frac{-ip\!\!\!/_{i}+m}{2m}.
\end{equation}
This is what the three-point correlator weighted by
the topological charge looks like, and is what we use to extract
the $\cancel{CP}$ form factors.

\section{Comparison of different topological charge definitions}
In this study, we use overlap fermions as valence quarks.
The overlap Dirac operator $D_{{\rm ov}}$ satisfies the Ginsparg-Wilson
relation, which ensures the lattice version of chiral symmetry at
finite lattice spacing $a$. 
Moreover, since the modified quark field $\hat{\psi}=(1-1/2D_{{\rm ov}})\psi$ is used for the chirally regulated
current operators and interpolating fields,
the effective quark propagator is then $1/(D_{c}+m_q)$, where $m_q$
is the current quark mass and $D_{c}=D_{{\rm ov}}/(1-1/2D_{{\rm ov}})$
 anticommutes with $\gamma_{5}$, i.e., $\{D_{c},\gamma_{5}\}=0$
\citep{Liu:2002qu}. This is the same form as in the continuum and
the eigenvalues of $D_{c}$ are purely imaginary.
Actually, it has been
shown that all the current algebra is satisfied with overlap fermions at finite $a$. In particular, the anomalous Ward identity (AWI) has been proven by Peter Hasenfratz~\citep{Hasenfratz:2002rp} for $D_{{\rm ov}}$ with chiral axial vector current.
And we have also shown numerically \citep{Liang:2018pis} that
the normalization factor $Z_{A}$ obtained from the axial Ward identity
in the isovector case is the same (within error) as the one from the
AWI in the singlet case.

Geometrically, the $\theta$ term is related to the topological charge
of the gauge field $Q_{t}=\int d^{4}xq_{t}(x)\equiv \frac{g^{2}}{32\pi^{2}}\int d^{4}xF_{\mu\nu}^{E}(x)\tilde{F}^{E,\mu\nu}(x)$.
Usually, the $F\tilde{F}$ definition of the
topological charge with unsmeared gauge fields suffers from large
UV effects and cannot give integer total topological charge values
on the lattice (a review on this topic can be found in \citep{Muller-Preussker:2015daa}).
One way to solve the problem is to use the gradient flow to smooth
the gauge fields and to get renormalized topological charges \citep{Luscher:2010iy,Luscher:2011bx,Luscher:2013cpa}.
Since we are using a lattice chiral fermion, we have an alternative way to obtain
the topological charge. According to the Atiyah-Singer index theorem,
the topological charge equals the numerical difference between 
the left-handed zero-modes of
$D_{{\rm ov}}$ and the 
right-handed zero-modes, that is, $Q_{t}=n_{-}-n_{+}$,
which ensures integer topological charge
on each configuration with no additional renormalization.
This definition is theoretically the same as the definition from the
overlap Dirac operator
\begin{equation}
Q_{t}=\frac{1}{2}{\rm Tr}\left[\gamma_{5}D_{{\rm ov}}\right]=-{\rm Tr}\left[\gamma_{5}\left(1-\frac{D_{{\rm ov}}}{2}\right)\right],
\end{equation}
where the trace over all color, spin and space-time indices of $D_{{\rm ov}}$ can be estimated through noise sources. And this
$D_{{\rm ov}}$ definition can also be used to define the topological charge density $q_t(x)$.
The topological charge term is essential in the nEDM calculation
and the overlap definition reduces the
subtleties in the evaluation of the topological charges, which
is another benefit of using chiral fermions.

\begin{figure}[ht]
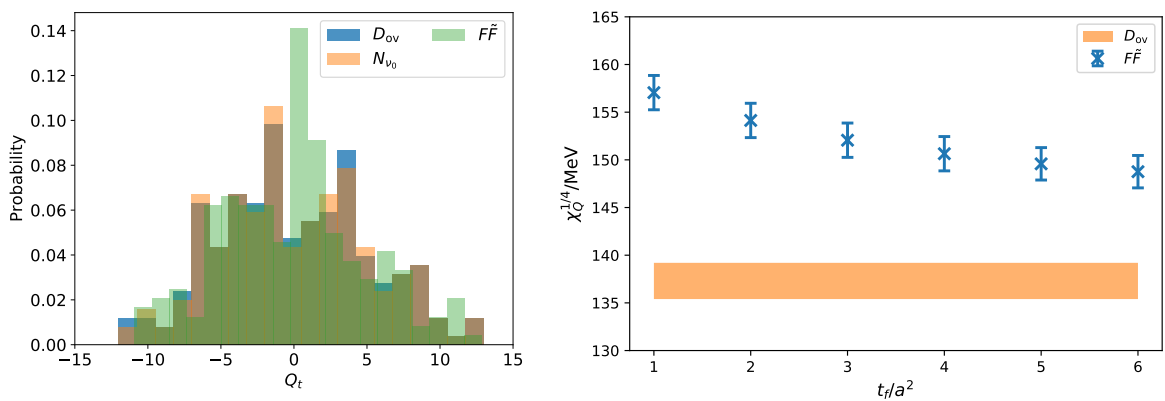

\begin{centering}
\includegraphics[width=0.40\textwidth,page=4, trim=0 0 0 0.2in]{figures/topo3}
\includegraphics[width=0.47\textwidth,page=3, trim=0 0.13in 0 0]{figures/Q2_005.pdf}
\par\end{centering}
\centering{}\caption{
Topological charge distributions over gauge configurations with different
definitions (left panel)
and the topological
susceptibility (right panel).
In the left panel,
the distribution with label $N_{\nu_{0}}$ corresponds
to the topological charges from counting the zero modes, which should
be the same as the one with label $D_{{\rm ov}}$. The nuanced difference
between them comes from the fact that $D_{{\rm ov}}$ is estimated
by noise and has statistical fluctuations. 
The distribution with label $F\tilde{F}$ corresponds to that using the gluonic definition with $t_f=4a^2$.
The brown color is the overlay of orange and blue. In the right panel,
the topological
susceptibility from the $F\tilde{F}$ definition is
plotted as a function of the flow time,
while the topological
susceptibility from the overlap definition is shown as a band.}
\label{fig:topological-distribution}
\end{figure}

It is interesting
to note the difference between topological charges from the overlap definition 
and those from the gluonic definition with long enough gradient flow until integer
topological charge values are reached. 
We find that, as shown in the left panel of Fig.~\ref{fig:topological-distribution}, 
the total topological charge on individual
configurations with the gluonic definition is not necessarily
the same as the one with the overlap definition. 
This is actually natural as they involve different regulations.
However, the topological charge
distributions over different gauge configurations in a given ensemble are similar. All the distributions
are approximately symmetric with central value at around zero, and
it seems the gluonic definition gives more zero charges. Now, a further question is whether they will lead to consistent
physical results at finite lattice spacing.

For the purpose of checking physical results, we calculate the topological
susceptibility on the same lattice
\begin{equation}
\chi_{t}=\frac{1}{V}\left\langle Q_{t}^2\right\rangle .
\end{equation}
The right panel of Fig.~\ref{fig:topological-distribution} shows that at large flow time
$t_{f}$, the value of the topological susceptibility from the gluonic
definition tends to approach that from the overlap definition. 
However, it is found that, even at $t_f=6a^2$,
the $\chi_t$ value from the gluonic definition 
is still around 10\% higher than that from the overlap definition although
there is a gentle trend that the central values will be
closer as the flow time
$t_f$ is larger still.
For the study at only one lattice spacing,
it is hard to justify a precise choice of $t_f$
that is large enough. On top of this, there is ${\cal O}(a^2)$ error.
Accordingly, in order to avoid such unnecessary systematic uncertainties, we use
the overlap definition of the topological charge in our calculation.
Another conclusion that can be drawn here is 
that the specific topological charge value
on each single configuration has not much effect on the physical correlations;
only the distribution matters.

\section{Data analysis details and systematics}

\subsection{Extracting Form Factors}

To calculate the $\cancel{CP}$ form factor, we need to the make three-point 
function to two-point function ratios

\begin{equation}
R_{3}\left(\Gamma_i,J_{\mu}\right)\equiv\frac{{\rm Tr}\left[\Gamma_i G_{3}\left(J_{\mu}\right)\right]}{{\rm Tr}\left[\Gamma_{e}G_{2}\right]}e^{E_{f}\left(t_f-t_c\right)}e^{E_{i}\left(t_c-t_0\right)},\label{eq:R01}
\end{equation}
and
\begin{equation}
R_{3}^{Q}\left(\Gamma_i,J_{\mu}\right)\equiv\frac{{\rm Tr}\left[\Gamma_i G_{3}^{Q}\left(J_{\mu}\right)\right]}{{\rm Tr}\left[\Gamma_{e}G_{2}\right]}e^{E_{f}\left(t_f-t_c\right)}e^{E_{i}\left(t_c-t_0\right)},\label{eq:R02}
\end{equation}
where $\Gamma_i$ is the polarized projector and $J_{\mu}$ stands
for the current insertion. If we write down the explicit form of 
the correlators, we have, e.g., in the $CP$ even case, 
\begin{equation}
R_{3}\left(\Gamma_i,J_{\mu};\vec{p}_{i},\vec{p}_{f},\vec{p}\right)=\frac{\frac{m^{2}}{E_{f}E_{i}}{\rm Tr}\left[\Gamma_i u(p_{f})\langle N_{f}|J_{\mu}|N_{i}\rangle\bar{u}(p_{i})\right]}{\frac{m}{E}{\rm Tr}\left[\Gamma_{e}u(p)\bar{u}(p)\right]}.\label{eq:R}
\end{equation}
Here again we assume $t$ is large enough to simplify
the equations. Details of dealing with the excited-states contamination
are discussed in the systematic uncertainty section. 
The additional overlapping and 
kinematic factors in Eqs.~(\ref{eq:R01}, \ref{eq:R02}) 
are cancelled with 
proper combination of two-point correlation functions.
Please note that in our numerical
setup we always set the initial momentum $\vec{p_{i}}=0$ in three-point
functions. 
With proper selection of the momentum $\vec{p}_{f}$, 
polarization $\Gamma_i$ and current insertion $J_{\mu}$, 
the ratio gives the desired nucleon matrix element for
particular form factors (or combinations of form factors). The relation
between the corresponding form factors and the setup of the ratios
are derived as follows.

For the normal EM case, we choose unpolarized projection and vector
current $\gamma_{4}$, which gives (in our momentum setup)

\begin{align}
 & R_{3}^{{\rm {\rm EM1}}}\left(\Gamma_{e},\gamma_{4}\right)\nonumber \\
= & \frac{{\rm Tr}\left[\Gamma_{e}G_{3}\left(\gamma_{4}\right)\right]}{{\rm Tr}\left[\Gamma_{e}G_{2}\left(\vec{p}=0\right)\right]}\nonumber \\
= & \frac{E_{f}+m}{2E_{f}}\left[F_{1}-\frac{|\vec{q}|^{2}}{2m\left(E_{f}+m\right)}F_{2}\right]\nonumber \\
= & \frac{E_{f}+m}{2{E_f}}G_{E},\nonumber 
\end{align}
where $G_{E}\equiv F_{1}-\frac{q^{2}}{4m^{2}}F_{2}$ is the electric
form factor. The last step used the fact that the momentum transfer
\begin{equation}
q^{2}=|\vec{q}|^{2}-\Delta E^{2}=|\vec{q}|^{2}-\left(E_{f}-m\right)^{2}
\end{equation}
and
\begin{equation}
|\vec{q}|^{2}=E_{f}^{2}-m^{2}.
\end{equation}
Therefore we have
\begin{equation}
q^{2}=E_{f}^{2}-m^{2}-\left(E_{f}-m\right)^{2}=\frac{2m}{E_{f}+m}|\vec{q}|^{2},
\end{equation}
and
\begin{equation}
\frac{|\vec{q}|^{2}}{2m\left(E_{f}+m\right)}=\frac{1}{2m\left(E_{f}+m\right)}\frac{E_{f}+m}{2m}q^{2}=\frac{q^{2}}{4m^{2}}.
\end{equation}

We can also choose polarized projection ($\Gamma_{i}\equiv\text{\ensuremath{-i\frac{1+\gamma_{4}}{2}}}\gamma_{5}\gamma_{i}$)
and the $\gamma_{j}$ current: 
\begin{align}
 & R_{3}^{{\rm EM2}}\left(\Gamma_{i},\gamma_{j}\right)\nonumber \\
= & \frac{{\rm Tr}\left[\Gamma_{i}G_{3}\left(\gamma_{j}\right)\right]}{{\rm Tr}\left[\Gamma_{e}G_{2}\left(\vec{p}=0\right)\right]}\nonumber \\
= & -\epsilon_{ijk}\frac{p_{f,k}}{2E_{f}}\left(F_{1}+F_{2}\right)\\
= & -\epsilon_{ijk}\frac{p_{f,k}}{2E_{f}}G_{M},\nonumber 
\end{align}
where $G_{M}\equiv F_{1}+F_{2}$ is the magnetic form factor, or unpolarized
projection and $\gamma_{i}$:
\begin{align}
 & R_{3}^{{\rm EM3}}\left(\Gamma_{e},\gamma_{i}\right)=\nonumber \\
= & \frac{{\rm Tr}\left[\Gamma_{e}G_{3}\left(\gamma_{i}\right)\right]}{{\rm Tr}\left[\Gamma_{e}G_{2}\left(\vec{p}=0\right)\right]}\nonumber \\
= & -i\frac{p_{f,i}}{2E_{f}}\left(F_{1}-\frac{q^{2}}{4m^{2}}F_{2}\right)\nonumber \\
= & -i\frac{p_{f,i}}{2E_{f}}G_{E}.\nonumber 
\end{align}
These ratios can be used to extract the $CP$ conserved form factors.
For the $\cancel{CP}$ case, we can choose the polarized projection
and $\gamma_{4}$, which turns out to be 
\begin{align}
R_{3}^{Q,{\rm EM}1}\left(\Gamma_{i},\gamma_{4}\right) & =\frac{{\rm Tr}\left[\Gamma_{i}G_{3}^{Q}\left(\gamma_{4}\right)\right]}{{\rm Tr}\left[\Gamma_{e}G_{2}\left(\vec{p}=0\right)\right]}\nonumber \\
 & =\frac{p_{f,i}}{2E_{f}}\left[\alpha^{1}F_{1}+\frac{E_{f}+3m}{2m}\alpha^{1}F_{2}+\frac{E_{f}+m}{2m}F_{3}'\right]\\
 & =\frac{p_{f,i}}{2E_{f}}\left[\alpha^{1}F_{1}-\frac{E_{f}-m}{2m}\alpha^{1}F_{2}+\frac{E_{f}+m}{2m}\left(2\alpha^{1}F_{2}+F_{3}'\right)\right]\nonumber \\
 & =\frac{p_{f,i}}{2E_{f}}\left[\alpha^{1}G_{E}+\frac{E_{f}+m}{2m}F_{3}\right].\nonumber 
\end{align}
An important fact about this ratio is that the neutron
form factor $F_{3,n}\left(0\right)$ has no $\alpha^{1}$ dependence
since $G_{E,n}\left(0\right)=0$. This means that one needs no
information about $\alpha^{1}$ or the other CP-even form factors if
one focuses only on the neutron case. Similarly, we can also use 
\begin{align}
R_{3}^{Q,{\rm EM}2}\left(\Gamma_{i},\gamma_{i}\right) & =\frac{{\rm Tr}\left[\Gamma_{i}G_{3}^{Q}\left(\gamma_{i}\right)\right]}{{\rm Tr}\left[\Gamma_{e}G_{2}\left(\vec{p}=0\right)\right]}\nonumber \\
 & =-i\left[\alpha^{1}\frac{E_{f}-m}{2E_{f}}\left(F_{1}+F_{2}\right)+\frac{p_{f,i}^{2}}{4mE_{f}}\left(\alpha^{1}F_{2}+F_{3}'\right)\right]\nonumber \\
 & =-i\left[\alpha^{1}\frac{E_{f}-m}{2E_{f}}G_{M}+\frac{p_{f,i}^{2}}{4mE_{f}}\left(\alpha^{1}F_{2}+F_{3}'\right)\right]\nonumber 
\end{align}
and 
\begin{align}
R_{3}^{Q,{\rm EM}3}\left(\Gamma_{i},\gamma_{j}\right) & =\frac{{\rm Tr}\left[\Gamma_{i}G_{3}^{Q}\left(\gamma_{j}\right)\right]}{{\rm Tr}\left[\Gamma_{e}G_{2}\left(\vec{p}=0\right)\right]}\nonumber \\
 & =-\frac{i}{4}\left[\alpha^{1}\frac{p_{f,i}p_{f,j}}{mE_{f}}F_{2}+\frac{p_{f,i}p_{f,j}}{{mE_{f}}}F_{3}'\right]=-\frac{i}{4}\frac{p_{f,i}p_{f,j}}{mE_{f}}\left[\alpha^{1}F_{2}+F_{3}'\right],
\end{align}
which prefers giving the combination of $\alpha^{1}F_{2}+F_{3}'$
rather than $F_{3}=2\alpha^{1}F_{2}+F_{3}'$. 

We use the ratios 
$R_{3}^{{\rm {\rm EM1}}}\left(\Gamma_{e},\gamma_{4}\right)$,
$R_{3}^{{\rm EM2}}\left(\Gamma_{i},\gamma_{j}\right)$,
and 
$R_{3}^{Q,{\rm EM}1}\left(\Gamma_{i},\gamma_{4}\right)$
in our calculation. 

\subsection{Summary on the systematic uncertainties}

In this study, the main sources of systematic uncertainties are
the two-state fits of the three-point (four-point) function to two-point function
ratios,
the momentum extrapolation, the use of the CDER technique, the final chiral extrapolation,
and the finite lattice spacing effect.

\begin{figure}
\begin{centering}
\includegraphics[page=1,width=0.45\textwidth]{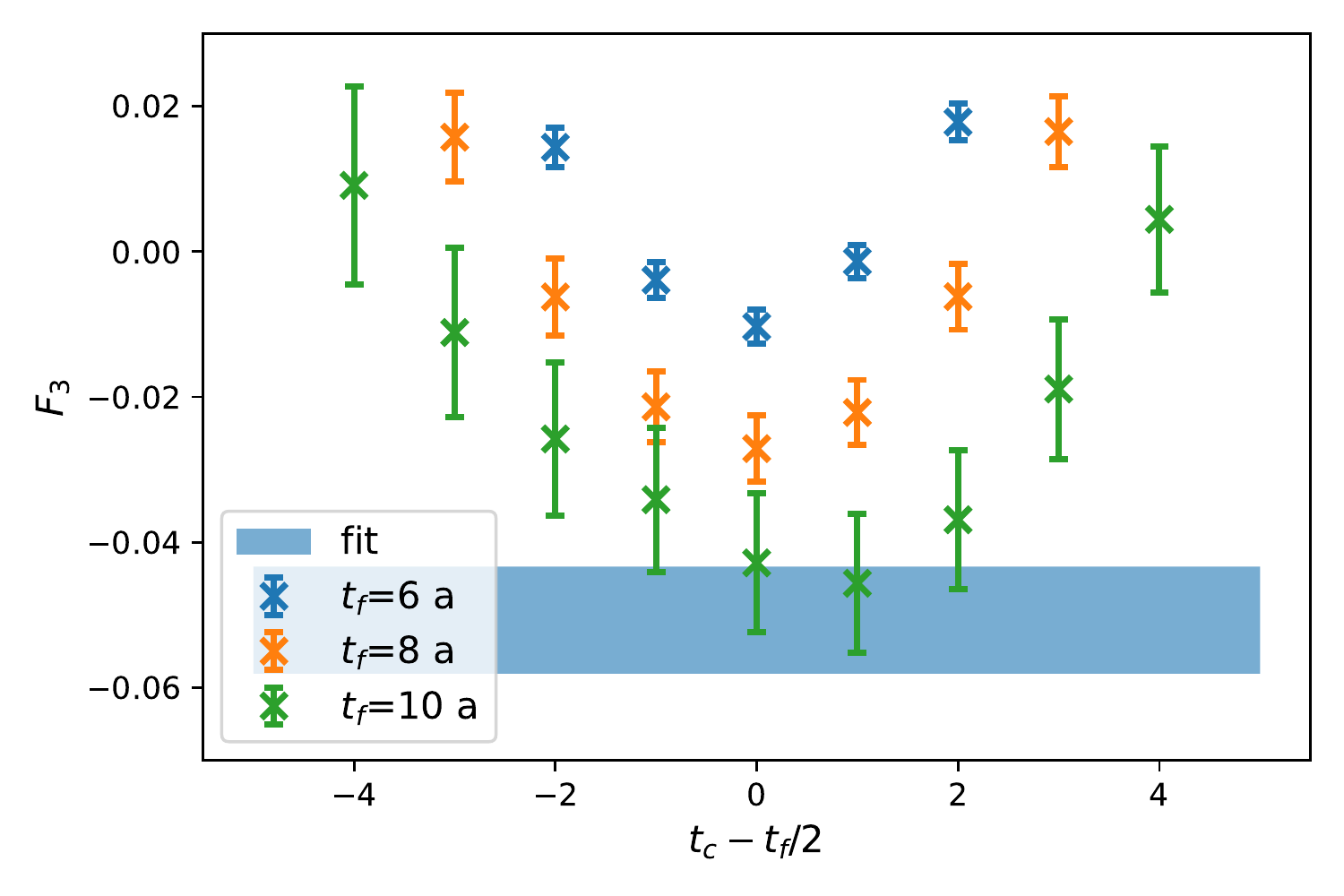}
\includegraphics[page=4,width=0.45\textwidth]{figures/EDM_005.pdf}
\par\end{centering}
\centering{\caption{An example of a two-state fit of $F_{3,n}$ with $Q^2=0.2$ GeV$^2$, $R=9a$ and $m_{\pi,v}\sim m_{\pi,s}=339$ MeV (left panel)
and the systematic uncertainty distribution over different momentum transfers, CDER cutoffs and pion masses (right panel).
\label{fig:2st}}}
\end{figure}

1) Two-state fit: The systematic uncertainty from the two-state fit
is estimated by the difference between
the two-state fitted values and 
the results from single-exponential fits using only the middle point at different separations.
Usually, one compares the two-state fits results and the values of the middle data point at the largest separations
to estimate the systematic uncertainty. In our case, since we are using relatively small source-sink separations,
we fit the middle points to a simplified form $C_0+C_1 e^{-mt_f}$
to account for the excited-state effect on different separations $t_f$.
Then, we consider the 
distribution of the difference between the two-state results and $C_0$'s
(as shown in the right panel of Fig.~\ref{fig:2st}),
and take the 1 $\sigma$ width (68\% probability) to be the 
the final systematic uncertainty,
which is determined to be \textbf{13\%}.

2) Momentum extrapolation: Considering the systematic uncertainty from the momentum extrapolation, 
although we have 5 momentum transfers,
the data points show no significant deviation from a
linear shape due to the large uncertainties,
so we use a linear fit for the extrapolation
and estimate the corresponding systematic uncertainty
to be the difference between the extrapolated value and
the data value with the smallest momentum transfer.
An example plot can be found in Fig.~\ref{fig:p-extra-sm}.
Similar to the two-state fit case,
the systematic uncertainty is estimated to 
be \textbf{10\%} by taking the 1 $\sigma$ width of the error distribution shown in the right panel of Fig.~\ref{fig:p-extra-sm}.
A fit with an additional $Q^4$ term results in no significant difference.

\begin{figure}
\begin{centering}
\includegraphics[page=5,width=0.45\textwidth]{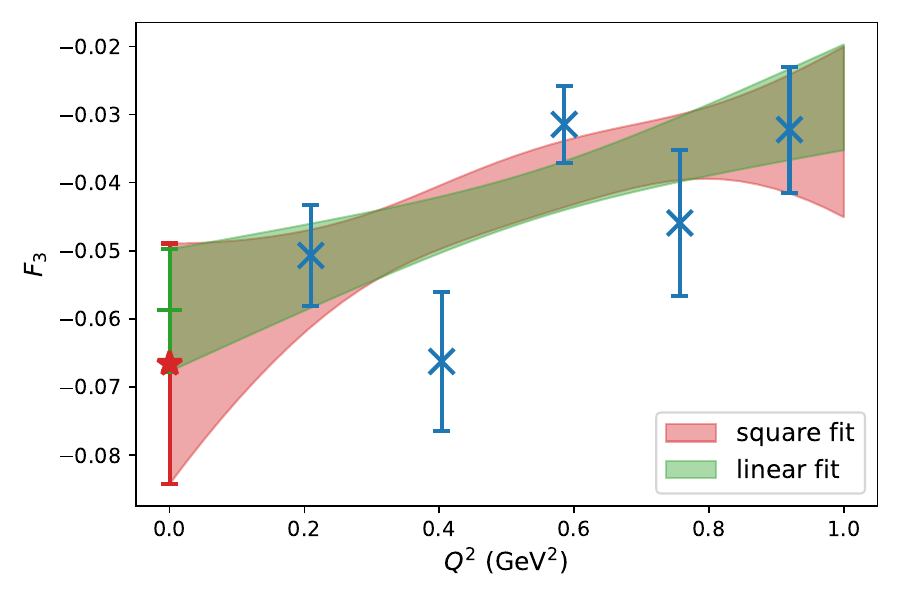}
\includegraphics[page=10,width=0.45\textwidth]{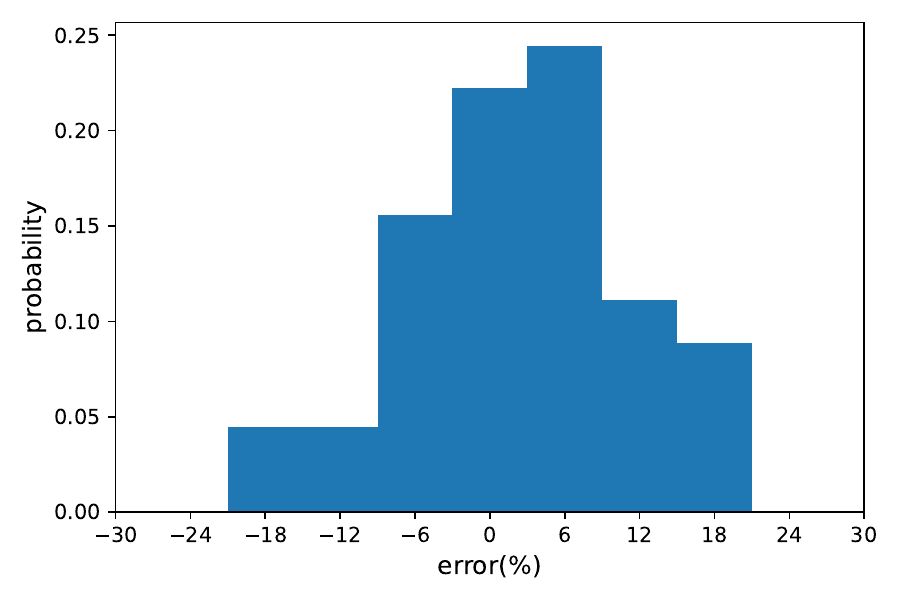}
\par\end{centering}
\centering{\caption{
An example of momentum transfer extrapolation of $F_{3,n}$ with $R=9a$ and $m_{\pi,v}\sim m_{\pi,s}=339$ MeV
(left panel)
and the systematic uncertainty distribution over CDER cutoffs and pion masses (right panel).
In the left panel,
blue points are lattice data and 
The green band shows a linear fit in $Q^2$ while the red band shows the fit with an additional $Q^4$ term.
\label{fig:p-extra-sm}}}
\end{figure}

\begin{figure}
\begin{centering}
\includegraphics[page=1,width=0.45\textwidth]{figures/EDM_all.pdf}
\includegraphics[page=2,width=0.45\textwidth]{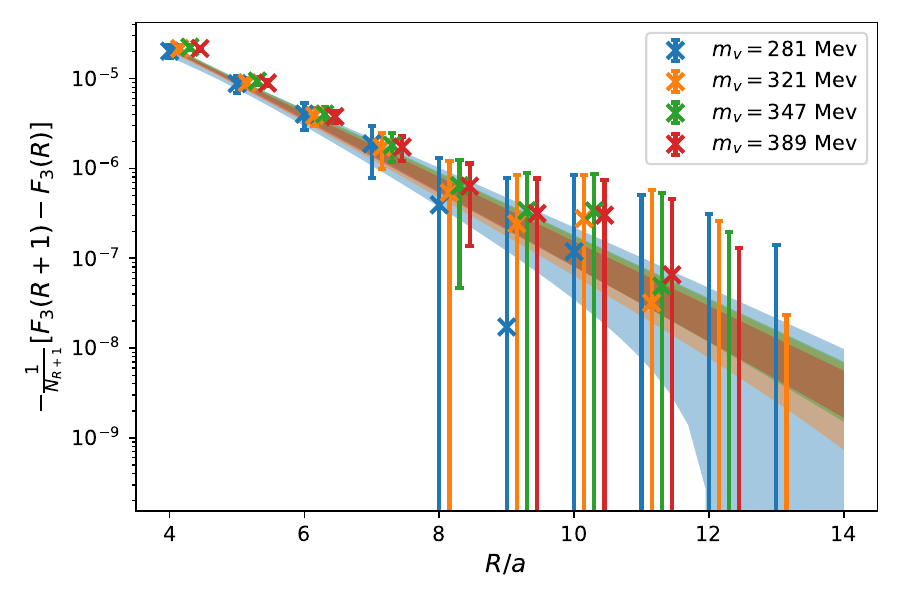}
\par\end{centering}
\centering{\caption{The left panel
shows the cutoff dependence of $F_{3,n}(Q^2=0.2~\mathrm{GeV}^2)$ with different $m_{\pi,v}$ and $m_{\pi,s}=339$~MeV,
while the right panel
shows the correlation in terms of the 
4-D distance $r$ between the topological charge operator and the
current operator.
Different colors are for different pion masses.
\label{fig:cder_error}}}
\end{figure}

\begin{figure}
\begin{centering}
\includegraphics[scale=0.48,page=3]{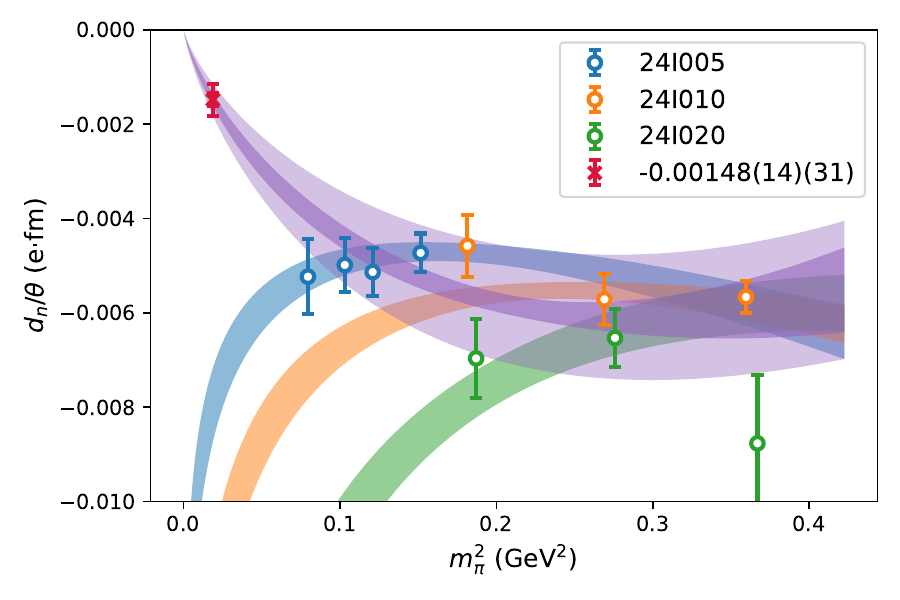}
\includegraphics[scale=0.48,page=4]{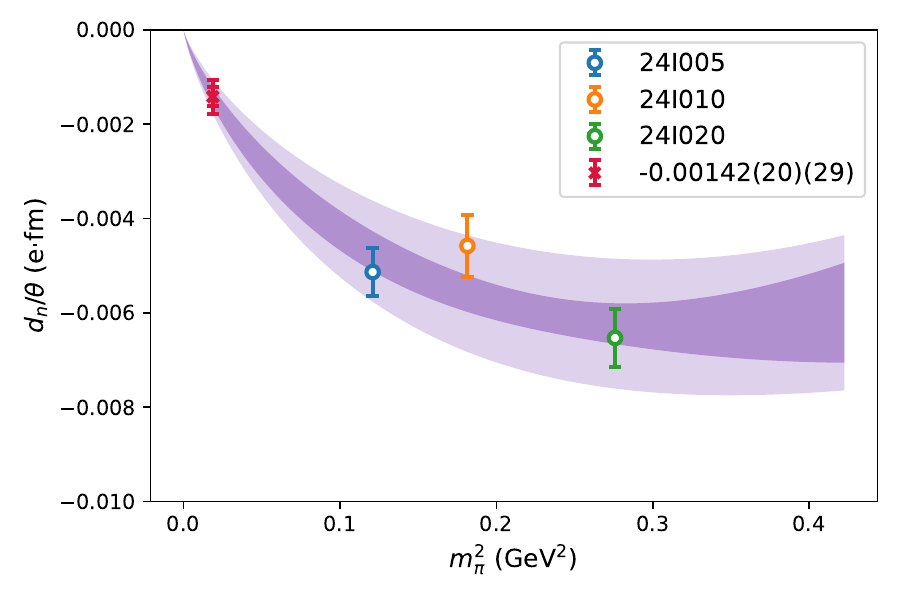}
\par\end{centering}
\centering{}\caption{
The chiral extrapolation of $d_n/{\theta}$ on both the sea and valence quark masses (left panel) 
and on only the unitary points (right panel).\label{fig:chiral_n}}
\end{figure}

\begin{figure}
\begin{centering}
\includegraphics[page=4,width=0.45\textwidth]{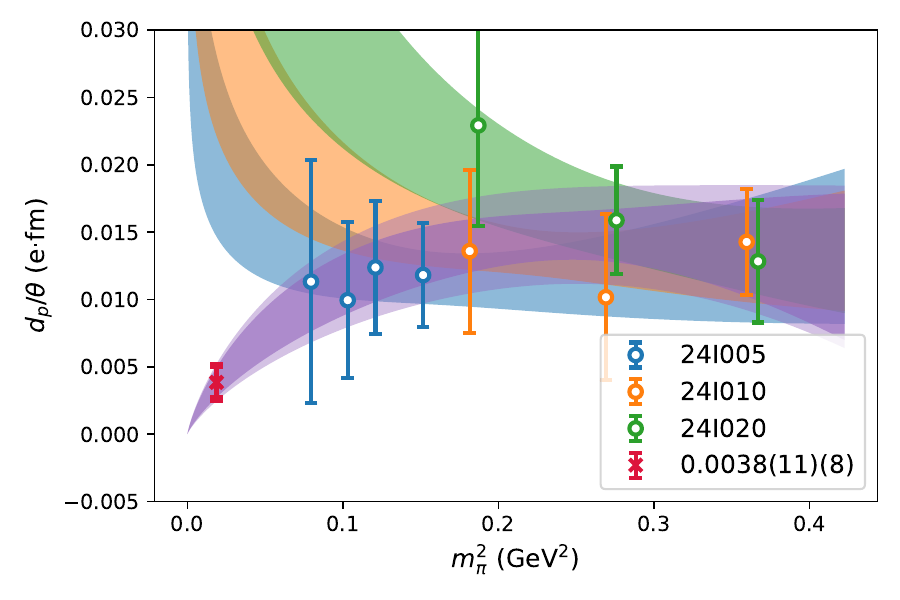}
\includegraphics[page=8,width=0.45\textwidth]{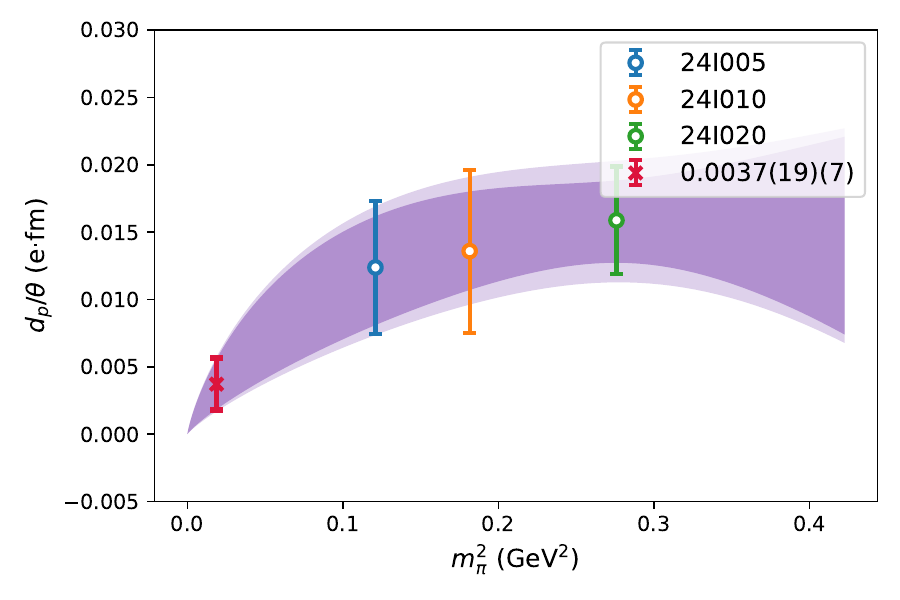}
\par\end{centering}
\centering{\caption{
The same as Fig.~\ref{fig:chiral_n} but for the proton case.
\label{fig:chiral_p}}}
\end{figure}

3) CDER technique: The systematic uncertainty due to the use of the CDER technique
is a crucial one.
The key idea of CDER is that
operators have finite correlation length and 
going beyond the correlation length
results in only noise rather than signal.
In our case, the topological charge operator is
summed up to a cutoff $R$ with the center being at
the position of the EM current.
We can have an optimal cutoff 
to have saturated signal and improved statistical error.
The left panel of Fig.~\ref{fig:cder_error}
shows the $d_n$ dependence on the cutoff $R$. We do observe
that, after $R\ge 9a$, the central values do not change (within errors)
while the errors are getting larger.
The right panel of Fig.~\ref{fig:cder_error}
shows the difference of $d_n$ normalized by the number of equivalent $R$'s
\begin{equation}
\frac{1}{N_{R+1}}\left[d_n(R+1)-d_n(R)\right],
\end{equation}
which is
in fact the correlation in terms of the 
4-D distance $r$ between the topological charge operator and the
current operator,
since 
\begin{equation}
d_n(R)\sim \sum_{|r|<R} \left\langle N| q(x+r) J_\mu(x) |N'\right\rangle,
\end{equation}
where $\left\langle N| q(x+r) J_\mu(x) |N' \right\rangle$ denotes the
nucleon matrix element that encodes the correlation.
This panel demonstrates that the 
correlation decays exponentially and there is  indeed a finite correlation length.
The optimal cutoff is chosen to be $R_0=9a$.
The systematic error can obtained by two ways. One is 
to take the difference between the value at 
$R_0=9a$ and the constant fitted value after that cutoff.
From data such as that in the left panel the systematic uncertainty is estimated to be $\sim$10--15\% in this way.
The other way is to fit the correlation to an exponential form first,
and then put the fitted correlation in the summation 
$d_n(R)\sim \sum_{|r|>R_0} \left\langle q(0+r) J_\mu(0) \right\rangle$
to estimate the contribution from the truncated tail.
In this way, with the correlation data such as that in the
right panel, the corresponding systematic uncertainty is estimated to be
$\sim$10\%.
So the two methods give consistent systematic uncertainties and
we choose \textbf{$\sim$12\%} to be our final estimation.

4) Chiral extrapolation: For the systematic uncertainty from the chiral extrapolation, we 
take the difference of the extrapolations with and without partially quenched data
points to be our estimation.
As shown in Fig.~\ref{fig:chiral_n} and Fig.~\ref{fig:chiral_p} (the chiral fits for proton),
the difference is around \textbf{3\%}.
The small systematic uncertainty of chiral interpolation is 
understandable since the chiral limit provides a very strong constraint
to the interpolation.

The total systematic uncertainty is found to be \textbf{21\%}, 
which is simply calculated by quadrature from all the systematic uncertainties.

\end{widetext}
\end{document}